\documentclass[useAMS,usenatbib,usegraphicx]{mn2e}
\usepackage{astroshortcuts} 
\renewcommand{\vec}[1]{\mathbfit{#1}}

\author[M. Zemp et al.]{Marcel Zemp$^1$\thanks{mzemp@phys.ethz.ch}, Joachim Stadel$^2$, Ben Moore$^2$ \& C. Marcella Carollo$^1$ \\
$^1$ ETH Zurich, Institute of Astronomy, Wolfgang-Pauli-Strasse 16, CH-8093 Zurich, Switzerland\\
$^2$ University Zurich, Institute for Theoretical Physics, Winterthurerstrasse 190, CH-8057 Zurich, Switzerland}

\title[Time-stepping scheme for $N$-body simulations]{An optimum time-stepping scheme for $N$-body simulations}

\begin{document}

\pagerange{\pageref{firstpage}--\pageref{lastpage}} \pubyear{2007}

\maketitle

\label{firstpage} 

\begin{abstract}
We present a new time-stepping criterion for $N$-body simulations that is based on the true dynamical time of a particle. This allows us to follow the orbits of particles correctly in all environments since it has better adaptivity than previous time-stepping criteria used in $N$-body simulations. Furthermore, it requires far fewer force evaluations in low density regions of the simulation and has no dependence on artificial parameters such as, for example, the softening length. This can be orders of magnitude faster than conventional ad-hoc methods that employ combinations of acceleration and softening and is ideally suited for hard problems, such as obtaining the correct dynamics in the very central regions of dark matter haloes. We also derive an eccentricity correction for a general leapfrog integration scheme that can follow gravitational scattering events for orbits with eccentricity $e\rightarrow1$ with high precision. These new approaches allow us to study a range of problems in collisionless and collisional dynamics from few body problems to cosmological structure formation. We present tests of the time-stepping scheme in $N$-body simulations of 2-body orbits with eccentricity $e\rightarrow1$ (elliptic and hyperbolic), equilibrium haloes and a hierarchical cosmological structure formation run.
\end{abstract}


\section{Introduction}

Achieving spatial adaptivity in the evaluation of forces in $N$-body simulations is a well-studied problem with many effective approaches based on the use of tree structures and multipole expansions or nested grids and FFT techniques. Such adaptivity in space also comes with a desire to achieve adaptivity in the time integration of these simulations since a large dynamic range in density implies a large dynamic range in time-scales for self gravitating systems ($T \sim 1/\sqrt{G \rho}~$). Two very different problems present themselves when trying to achieve this. First, there are no practical (explicit), general purpose (applicable to a wide range of astrophysical problems), adaptive integration techniques known for the $N$-body problem which are symplectic. By this we mean that the numerical integration is an exact Hamiltonian phase flow very close to the phase flow of the continuous system under study. This is a very desirable property for following systems for longer than a single dynamical time. Such exact preservation of the geometrical properties of the dynamical system is possible for fixed time-step schemes. For general $N$-body simulations with adaptive time-stepping we have to resort to approximate symplectic behaviour or preservation of time symmetry (a property which is known to lead to very good integration methods). The second problem is that of continuously determining the appropriate time-step for each particle in the simulation so that the error in the integration remains within tolerance while performing the fewest possible force evaluations and minimising the computational overhead. Resolving this second problem is the main focus of this paper and can be considered independently from methods symmetrizing the time-stepping scheme such as presented in \cite{2001PhDT........21S} and \cite{2006NewA...12..124M}.

There are several known time-step criteria based on different properties of the simulation (e.g. local density $\rho(r)$, potential $\Phi(r)$, softening $\epsilon$, acceleration $a$, jerk $\dot{a}$ or even the velocity $v$ of the particle) that are used in state-of-the-art numerical codes \citep{1997astro.ph.10043Q, 2001PhDT........21S, 2003gnbs.book.....A, 2003MNRAS.338...14P, 2005MNRAS.364.1105S}. Some of them have a physical motivation, others are just a clever combination of physical properties in order to obtain a criterion which has the physical unit of time. All are an attempt to find an inexpensive way of determining an appropriate time-step for each particle in the simulation. For cosmological simulations, the ad-hoc time-step criterion based on the acceleration and the gravitational softening of the particle ($T \sim \sqrt{\epsilon / a}~$) has proven very successful, despite the fact that it is not directly related to the dynamical time in these simulations. One reason why this time-step criterion is thought to work well is that it results in a very tight time-step distribution with very infrequent changes in the time-step of a particle in block time-stepping schemes (power of 2 step sizes; see also section \ref{chap:implementation}). This hides the evils due to the first problem since the behaviour is more like that of a fixed time-stepping scheme than for other criteria. The price, however, is that more time-steps are taken in lower density regions of the simulation than would seem to be necessary. Furthermore, while still adequate for the type of simulations that have been performed up to now, which adapt the softening and mass of particles in the highest density regions and thus reduce the time-step somewhat artificially in those regions, this time-stepping scheme is no longer effective in simulations covering a much larger dynamic range. 

In state-of-the-art computer simulations, structures can be resolved by $\Nvir \approx O(10^7)$ which results in a resolution scale of $r_{\mathrm{res}} \approx O(10^{-3}~\rvir)$. By comparing the dynamical scales at $\rvir$ and $r_{\mathrm{res}}$, we get $T_{\mathrm{dyn}}(\rvir)/T_{\mathrm{dyn}}(r_{\mathrm{res}}) \approx O (10^3)$ or even larger for future high resolution simulations. This large dynamical range in time-scales demands an acutely adaptive criteria so that the dense, dynamical active regions are resolved correctly while the simulations remain fast.

However, in a general simulation, such as those used for cosmological structure formation, it is not straightforward to determine the dynamical time of a given particle. This depends on the dominant structure responsible for the orbit of a particle which needs to be quickly determined at each time-step as the particle is advanced. In this paper we develop a new fast method of determining each particle's true dynamical time using information directly computed in the force evaluation stage of the simulation. This is quite different from using the local dynamical time which fails dramatically under many circumstances, e.g. consider using the local density near the earth to estimate its time-step!

We present in Section \ref{chap:TS} the basic ideas and our implementation of the new adaptive dynamical time-stepping criterion into a tree-code. Detailed and extensive tests are presented in Section \ref{chap:TSTest}. In Section \ref{chap:conclusions} we conclude.

\section{Dynamical time-stepping} \label{chap:TS}

\subsection{General idea and description}

In order to advance a particle in a numerical simulation, we have to choose a particular time-step for each individual particle. Let us consider a particle on a circular orbit in a system with spherical symmetric density profile $\rho(r)$. The dynamical time $T_{\mathrm{dyn}}(r)$ (or orbital time) of this particle at radius $r$ under spherical symmetry is given by
\begin{equation}
T_{\mathrm{dyn}}(r) = 2 \pi \frac{1}{\sqrt{G \rho_{\mathrm{enc}}(r)}}~,
\end{equation}
where 
\begin{equation}
\rho_{\mathrm{enc}}(r) \equiv \frac{M(r)}{r^3}
\end{equation}
is the enclosed density within the radius $r$ and $M(r)$ is the total mass within radius $r$. The natural choice for a time-step of a particle would therefore be $\Delta T \propto T_{\mathrm{dyn}}$ were we not faced with the difficulty of determining the enclosed density.

For a particle in a given landscape of cosmic structure, the enclosed density should be set, roughly speaking, by the structure that the particle is orbiting about. Within collisionless cosmological simulations this could be some super-cluster scale structure, or an individual dark matter halo, or some substructure within a dark matter halo. Ideally, we would scan the whole sky of the particle and determine the structure that gives the dominant contribution to its acceleration. From this dominating structure, we could determine the enclosed density and hence find the dynamical time of the particle.

Here we have to distinguish two different regimes. First, we have the mean field regime, i.e. particles move in a (slowly) varying potential that is determined by the total mass distribution. The individual particles are only weakly influenced by their direct neighbours and their motions are dictated by the sum of more distant particles. This is ensured by appropriately softening the short range force thereby placing an upper limit on the contribution from an individual particle. In this regime we want the enclosed density to be set by the globally dominating structure. The second regime is the gravitational scattering regime where we would like to follow large angle scattering due to gravitational interactions, i.e. orbits with eccentricity $e\rightarrow1$. Here it is important to get the contributions from the closest neighbours which dictate the orbital evolution when they are very close and when there is little or no force softening. This means that the enclosed density is often set by some locally dominating particle.

The determination of the enclosed density is quite easy for some simple configurations like the 2-body problem or a particle orbiting an analytically given spherical symmetric structure. However, the generalisation to any given configuration in an $N$-body simulation is not so straightforward and we present a simple way in which this can be achieved within a tree based gravity code. The specific implementation within other code-types may look somewhat different but the general scheme and spirit of the method stay the same.

\subsection{Implementation within a tree-code} \label{chap:implementation}

We use the tree-code \textsc{pkdgrav} written by Joachim Stadel \citep{2001PhDT........21S} which allows for an adaptive time-stepping mechanism where each particle can be on a different time-step. The time-steps of the particles are quantized in fractions of powers of two of a basic time-step $T_0$ (block time-stepping). Therefore, particles on rung $n$ have an individual time-step of
\begin{equation}
\Delta T = \frac{T_0}{2^n}~,
\end{equation}
where $T_0$ is the basic time-step of the simulation and can be chosen by the simulator. As stated previously, our  time-stepping criterion is given by,
\begin{equation}\label{eq:TSdyn}
\Delta T_{\mathrm{D}} = \frac{T_0}{2^n} \leq \eta_{\mathrm{D}} \frac{1}{\sqrt{G \rho_{\mathrm{enc}}(r)}}~,
\end{equation}
where $\eta_{\mathrm{D}}$ is a free parameter. Therefore, we need to calculate the enclosed density $\rho_{\mathrm{enc}}$ for each particle from information that is available in a tree-code in order to determine its rung.

In hierarchical tree-codes, at every time-step two interaction lists are generated for each particle: a list of cells and a list of particles that interact with the given particle. The tree structure in such codes is a hierarchical representation of the mass in the simulation with each subvolume, or cell, being a node in this tree. As we proceed from the root to the leaves of this data structure we get an ever finer mass representation of the simulation.  The forces from more distant cells are calculated by using the multipole expansion of the gravitational potential. This expansion makes it clear that a finer mass representation, or smaller cell, is required for nearby regions than for more distant ones if we want uniform relative errors for each multipole contribution to the force. In its simplest form this is realised by a tree-walk algorithm which, for a given cell, decides whether the use of a multipole expansion for this cell satisfies a given error tolerance. If not, this cell is opened and its two or more children are considered in the same way. The opening radius of a cell which sets an error tolerance is defined by
\begin{equation}
r_{\mathrm{open}} = \frac{2}{\sqrt{3}} \frac{r_{\mathrm{max}}}{\theta}~,
\end{equation}
where $\theta$ is the opening angle and $r_{\mathrm{max}}$ is the distance from the centre of mass of the cell to the most distant corner of the cell. The numbers are only geometric factors so that in the case of a cubic cell with homogeneous density $2/\sqrt{3}~r_{\mathrm{max}}$ corresponds to the side length of the cube. A cell may only be accepted as a multipole interaction if the particle for which we are calculating the force is further from the centre of mass of the cell than this radius. If a leaf cell (called buckets) needs to be opened, then we calculate the interactions with each of its particles directly (no multipole expansion is used in this case). 

At the end of this procedure each particle in the simulation has two interaction lists: 1) a cell list which can be thought of as the long range contributions to gravity and 2) a particle list which accounts for the short range gravitational interactions. The acceleration and the potential energy of each particle are calculated from these two interaction lists. The opening angle varies the ratio of directly calculated forces to those calculated via multipole expansions. It therefore controls force errors and also determines the primary cost of a simulation. 

For the calculation of the dynamical time of a particle, we generate an additional cell list which provides a reduced representation of the particle list, i.e. this list only contains the buckets that were opened by the above procedure and these buckets are treated in the same way as the cells for the long range contributions. The cell list and this additional list, which we call the particle-bucket list, form a complete tiling of the entire simulation volume except for the local bucket of the particle itself, which is not included.

\subsubsection{Mean field regime algorithm} \label{chap:meanfield}

We only need to calculate the time-step of a particle when we evaluate the force acting on it. The dynamical time of a particle is then determined according to the following scheme:

\begin{enumerate}
\item Pick out the 0.5 percentile highest values of $\rho_{\mathrm{enc}}$ from both the cell and particle-bucket interaction lists. We regard this subset of cells as the centres of dominant contributions to the acceleration of the particle, otherwise called maxima. Once we have added the contributions of the mass surrounding each of these centres we can make a final determination of which is the dominating region and hence set the correct dynamical time-step. The enclosed density for a cell is defined by,
\begin{equation}
\rho_{\mathrm{enc}} = \frac{M_{\mathrm{C}}}{|\vec{r}_{\mathrm{PC}}|^3}~,
\end{equation}
where $M_{\mathrm{C}}$ is the total mass of the cell and $\vec{r}_{\mathrm{PC}}$ the vector from the location of the particle to the centre of mass of the cell.
\item For each of these centres, add up all the $\rho_{\mathrm{enc}}$ values from the other cells in the list that satisfy both of the following criteria:
\begin{eqnarray}
|\vec{r}_{\mathrm{PC}}| &\leq& 2~|\vec{r}_{\mathrm{PCmax}}|\\
0.75 &\leq& \frac{\vec{r}_{\mathrm{PCmax}} \cdot \vec{r}_{\mathrm{PC}}} {|\vec{r}_{\mathrm{PCmax}}| |\vec{r}_{\mathrm{PC}}|}~,
\end{eqnarray}
where $\vec{r}_{\mathrm{PC}}$ is the vector from the location of the particle to the centre of mass of the cell and $\vec{r}_{\mathrm{PCmax}}$ is the vector from the particle to one of the maxima. That means, we add up all the $\rho_{\mathrm{enc}}$ values of cells that lie within a spherical viewing cone of opening angle $2 \alpha = 2 \arccos(0.75) \approx 83^\circ$ around a maximum cell (C$_\mathrm{max}$) with the particle (P) being the apex extending to $2~|\vec{r}_\mathrm{{PCmax}}|$.\footnote{Adding up the $\rho_{\mathrm{enc}}$ values shows less scattering in the determined time-steps than adding up first the masses of the cells and then dividing by the total volume. It also correctly accounts for softened contributions to the force from the region close to P since the $\rho_{\mathrm{enc}}$ contributions are reduced there.} See also Fig. \ref{fig:cone} for the geometric configuration. If the particle would orbit a perfectly spherically symmetric halo at radius $r$ then the dynamical relevant mass would lie in the sphere of radius $r$ centred at the geometric centre of the halo. Therefore, the angle $\alpha$ is chosen so that the volume of the sphere
\begin{equation}
V_{\mathrm{S}} = \frac{4 \pi}{3} r^3
\end{equation}
equals the volume of the spherical cone
\begin{equation}
V_{\mathrm{C}} = \frac{2 \pi}{3} (2r)^3 [1 - \cos(\alpha)]
\end{equation}
resulting in
\begin{equation}
\frac{V_{\mathrm{S}}}{V_{\mathrm{C}}} = \frac{1}{4 [1 - \cos(\alpha)]} = 1~.
\end{equation}
This is reached for $\cos(\alpha) = 0.75$.

\begin{figure}
		\includegraphics[width=\columnwidth]{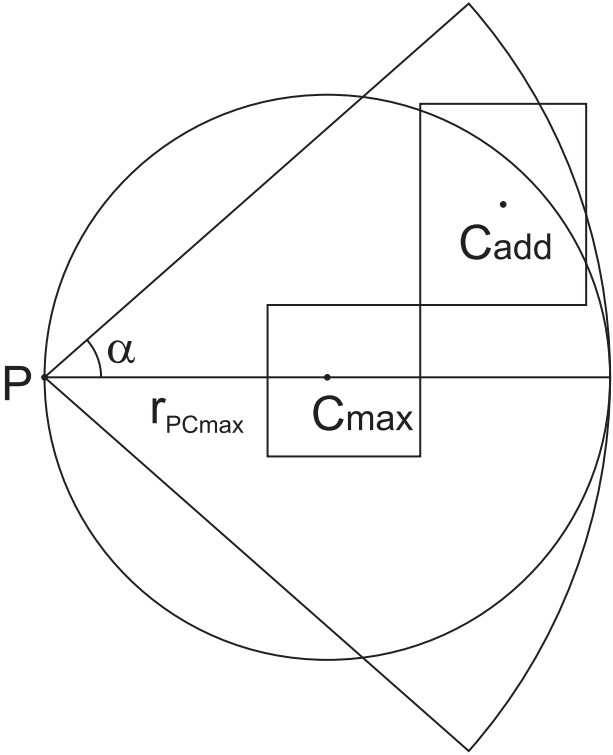}
		\caption{Viewing cone for the allowed region of cells to be accepted by the time-step criterion. C$_{\mathrm{max}}$ is the location of the maximum cell. We accept all cells that are within the cone of opening angle $2\alpha \approx 83^\circ$ with the particle (P) being the apex extending to $2~|\vec{r}_\mathrm{{PCmax}}|$.}
	\label{fig:cone}
\end{figure}

\item We now have a summed $\rho_{\mathrm{enc}}$ of mass contributions about each maximum. Only the largest of these, $\rho_{\mathrm{enc,MF}}$, is used in determining the dynamical time-step of the particle.

\item Add the local density $\rho_{\mathrm{local}}$ to the enclosed density $\rho_{\mathrm{enc,MF}}$ of the particle. We do this in order to account for possible contributions from the local bucket of the particle.
\end{enumerate}

As prefactor we use a value of $\eta_{\mathrm{D}} = 0.03$. This choice is motivated by earlier studies in \cite{2001PhDT........21S} and is further discussed in section \ref{chap:paramdep}.

\subsubsection{Error terms in leapfrog schemes}

In computer simulations, the system is not evolved by the true Hamiltonian but by the approximate Hamiltonian
\begin{equation}
H_{\mathrm{A}} = H_0 + \Delta T^2 H_2 + \Delta T^4 H_4 + O(\Delta T^6)~.
\end{equation} 
In \textsc{pkdgrav}, particles are evolved by using a kick-drift-kick leapfrog scheme. This means 
\begin{eqnarray}
H_0 &=& H_{\mathrm{D}} + H_{\mathrm{K}} = H \\
H_2 &=& \frac{1}{12} \left\{\left\{H_{\mathrm{K}},H_{\mathrm{D}}\right\},H_{\mathrm{D}}\right\}
- \frac{1}{24} \left\{\left\{H_{\mathrm{D}},H_{\mathrm{K}}\right\},H_{\mathrm{K}}\right\}
\end{eqnarray}
where we have split the true Hamiltonian $H$ into a drift ($H_{\mathrm{D}}$) and kick ($H_{\mathrm{K}}$) part. A detailed derivation and an expression for $H_4$ is given in the Appendix. Therefore, the dominant error term is the second order term $E_2 = \Delta T^2 H_2$.

For a 2-body problem, the Hamiltonian is given by
\begin{equation}
H = \underbrace{\frac{p_r^2}{2 \mu} + \frac{p_\varphi^2}{2 \mu r^2}}_{H_{\mathrm{D}}} 
\underbrace{- \frac{A}{r}}_{H_{\mathrm{K}}}
\end{equation}
where $\mu \equiv \frac{M_1 M_2}{M_1 + M_2}$ is the reduced mass and $A = G M_1 M_2$ and where $M_1$ and $M_2$ denote the masses of the two particles. The problem is described by the two coordinates $r$ and $\varphi$ and their conjugate momenta
\begin{eqnarray}
p_r &=& \mu \dot{r} \\
p_\varphi &=& \mu r^2 \dot{\varphi} = L
\end{eqnarray}
Since the coordinate $\varphi$ is cyclic, its conjugate momentum is an integral of motion, i.e., the angular momentum,
\begin{equation}
L^2 = \mu a A (1-e^2),
\end{equation}
is conserved. Here $a = \frac{A}{-2E}$, i.e. $|a|$ is the semimajor axis of the ellipse ($e<1$) or hyperbola ($e>1$) and $E$ is the total energy of the orbit. By using a symmetrised time-step,
\begin{equation} \label{eq:Tdynsym}
\Delta T = \eta_{\mathrm{D}} \sqrt{\frac{r^3}{G \left(M_1 + M_2\right)}} = \eta_{\mathrm{D}} \sqrt{\frac{r^3 \mu}{A}},
\end{equation}
we can calculate the higher order error term $E_2$ of the approximate Hamiltonian for a 2-body problem and evaluate it at pericentrer of the particle's orbit,
\begin{equation}
E_2^{\mathrm{peri}} = \Delta T^2 H_2 = \frac{1}{24} \frac{(1 + 2 e)~\eta^2_{\mathrm{D}} A}{(1-e)~a}.
\end{equation}

We see that the error depends on eccentricity $e$ of the orbit. This allows us to correct for the second order error and control the error at pericentre.

\subsubsection{Gravitational scattering regime algorithm}

If the interaction of a particle is in the gravitational scattering regime (e.g. it is located close to a super-massive black hole), we first determine the mean field value $\rho_{\mathrm{enc,MF}}$ for this particle in exactly the same way as described in section \ref{chap:meanfield}. However, in order to account for gravitational scattering events we need to consider the close particle interactions in our determination of the dynamical time in more detail. The procedure here is the following: we go through the particle interaction list and pick out the highest value of
\begin{equation}\label{eq:REPsym}
\rho_{\mathrm{enc,GS}} = C(e)~\frac{M_{\mathrm{P}} + M_{\mathrm{I}}}{|\vec{r}_{\mathrm{PP}}|^3},
\end{equation}
where $M_{\mathrm{P}}$ is the mass of the particle, $M_{\mathrm{I}}$ is the mass of the particle in the interaction list, $\vec{r}_{\mathrm{PP}}$ is the particle-particle distance and
\begin{equation}
C(e) \equiv \frac{1 + 2 e}{|1-e|}
\end{equation}
is the additional factor that corrects for eccentricity of the orbit. The symmetrisation in $\rho_{\mathrm{enc,GS}}$ is to cover cases where unequal mass particles are involved in the interaction. In such cases the heavier particle would be on a much larger time-step than the interacting partner, resulting in momentum conservation problems and unphysical behaviour when the mass ratio is large. The eccentricity of two interacting particles is given by
\begin{equation}
e \equiv \sqrt{1+\frac{2 E L^2}{\mu A^2}}~.
\end{equation}
Hence, for each particle we would like to follow in the gravitational scattering regime, we have calculated the two values of $\rho_{\mathrm{enc,MF}}$ and $\rho_{\mathrm{enc,GS}}$. The larger of these two is then used.

\section{Time-stepping criterion tests and behaviour}\label{chap:TSTest}

\subsection{General properties}

\begin{figure*}
		\includegraphics[width=0.495\textwidth]{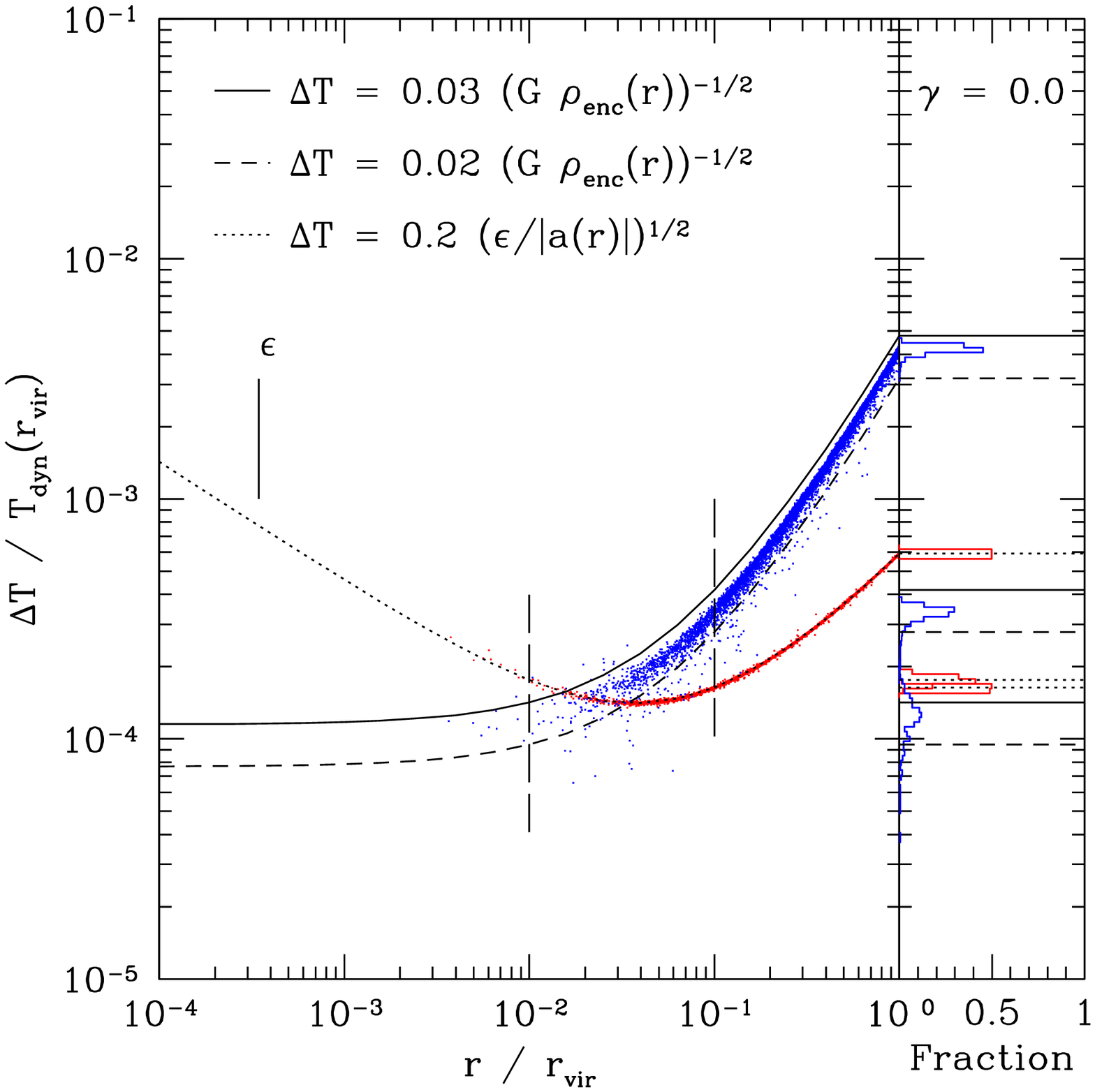}\hfill
		\includegraphics[width=0.495\textwidth]{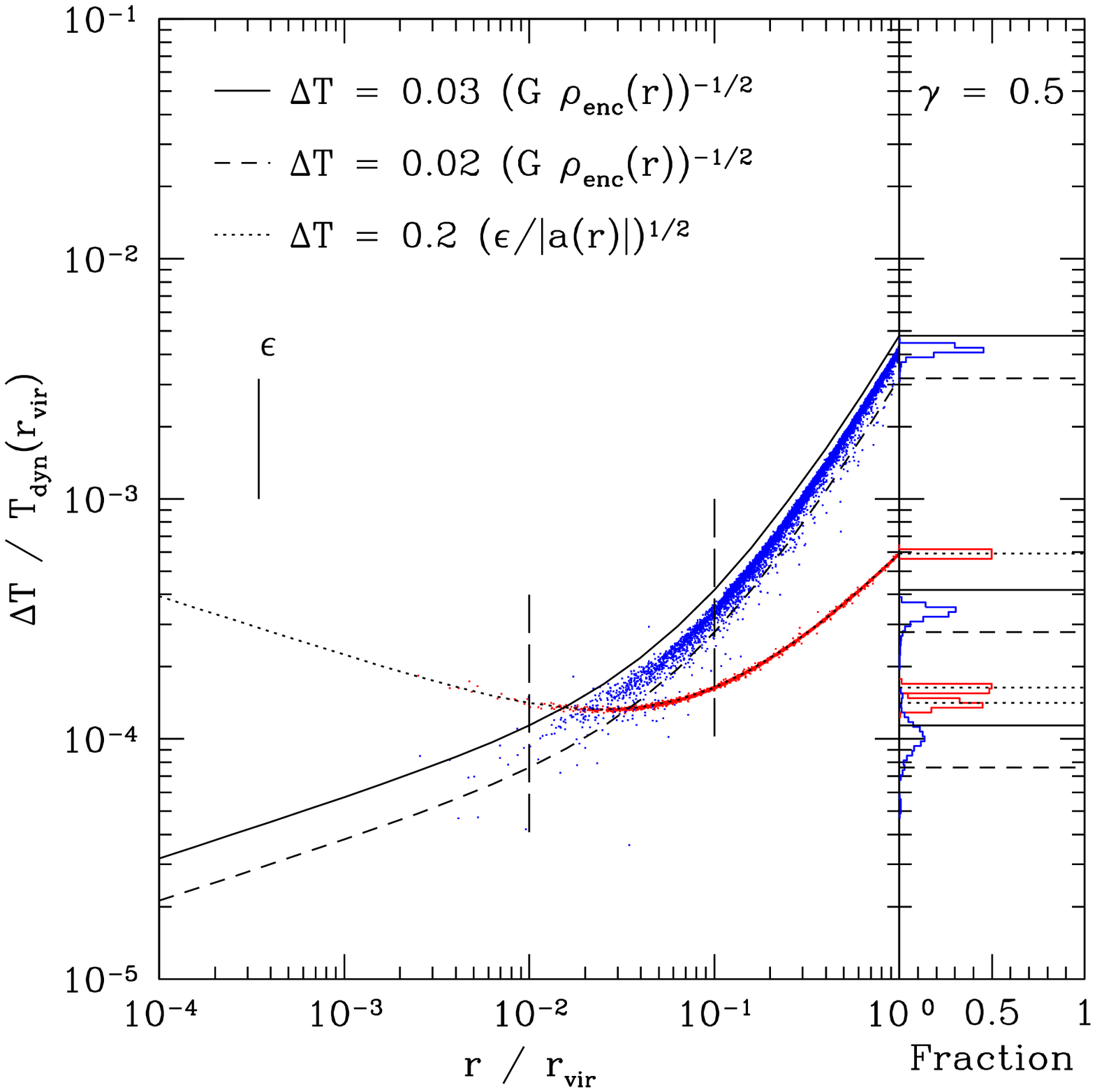}\\
		\includegraphics[width=0.495\textwidth]{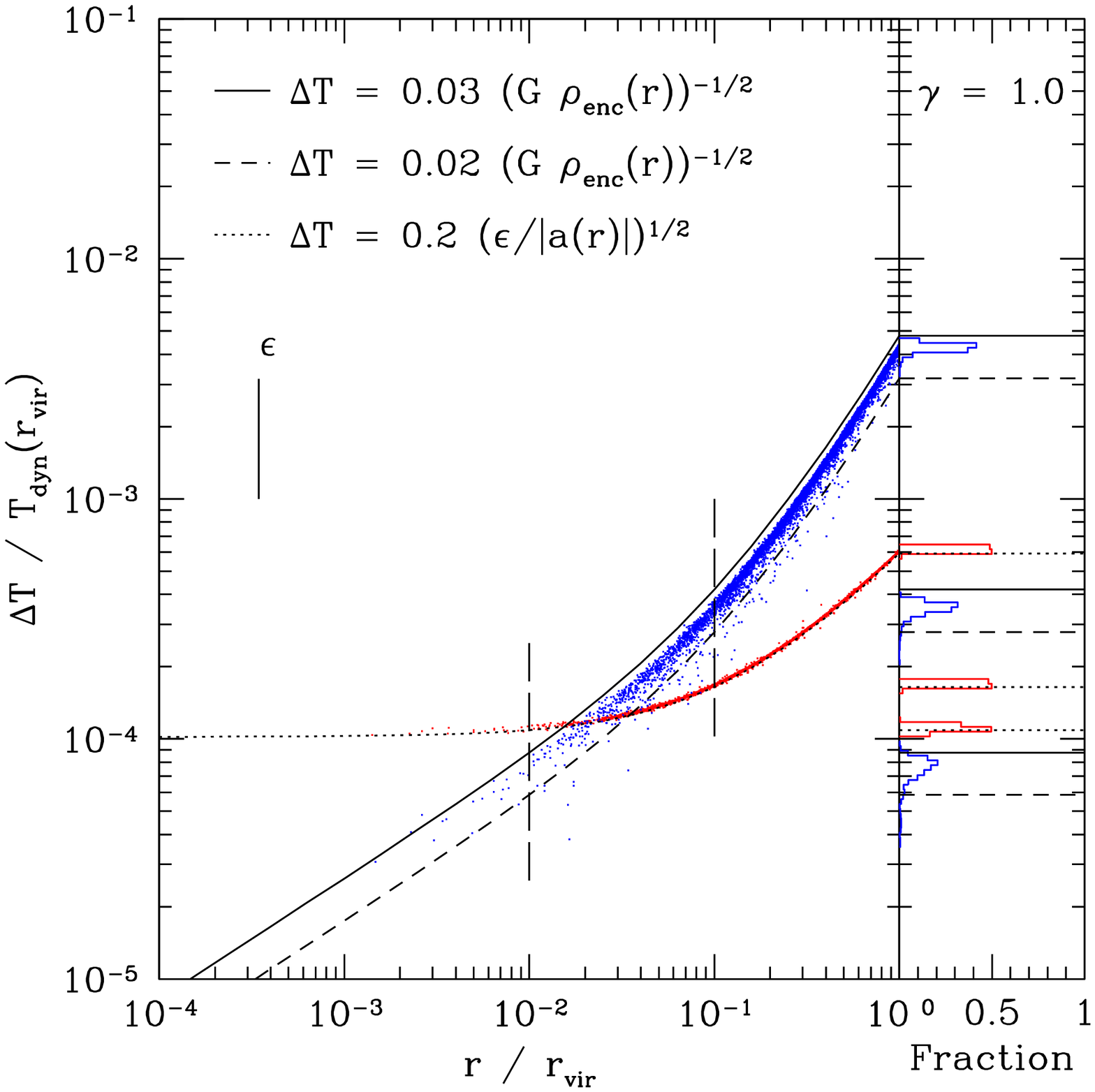}\hfill
		\includegraphics[width=0.495\textwidth]{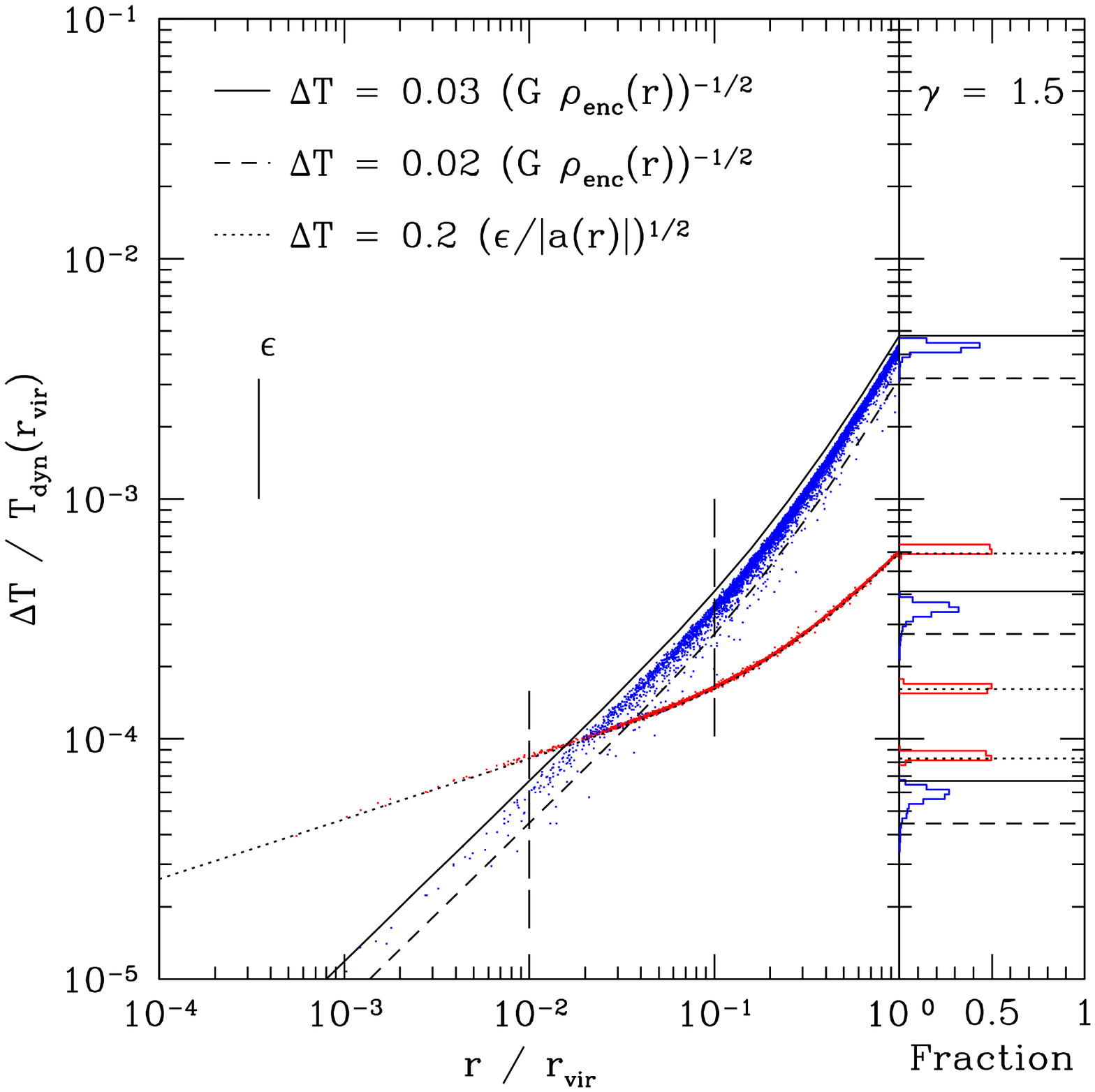}\\
		\caption{We plot the time-step criterion distribution for four profiles with central slope ranging from $\gamma = 0.0~...~1.5$. The results for the dynamical and standard time-stepping criteria are shown - between solid and long dashed lines: dynamical time-step criterion (blue) and following the short dashed line: standard time-step criterion (red). Time-steps are in units of the dynamical time at the virial radius $T_{\mathrm{dyn}}(\rvir) \approx 12~\mathrm{Gyr}$. Additionally, we plot the theoretical curves for the standard and dynamical time-stepping criterion we expect for a spherical symmetric halo with the given profile. It is evident that the dynamical criterion is much more adaptive with radius than the standard criterion. The time-steps taken using the standard criterion remain constant over larger spans in radius than with the dynamical time-step criterion. For example, in the NFW profile ($\gamma = 1.0$), all particles with a distance smaller than $\approx 0.1~\rvir$ from the centre are on a single constant time-step. Both criteria give equal time-steps at the radius $r_{\mathrm{eq}} = 4.444~\mathrm{kpc} \approx 1.5 \times 10^{-2}~\rvir$ - independent of the density profile.	For flat central profiles, the time-step increases below the radius where the acceleration has its maximum! One can also see the asymptotic radial behaviour of both schemes in the central region given by equations (\ref{eq:tstdbeh}) respectively (\ref{eq:tdynbeh}). We also plot the time-step distribution of the dynamical time-step criterion in three bins at $r_i = \rvir, 10^{-1}~\rvir, 10^{-2}~\rvir$ of width 0.002 in logarithmic scale on the right side of each plot. We can see that the dynamical time-step criterion follows closely a band between $\eta_{\mathrm{D}} = 0.02 ... 0.03$ (long dashed and solid lines). In the two flat cases ($\gamma = 0.0$ and $\gamma = 0.5$), the bin at $10^{-2}~\rvir$ is close to the resolution limit and the distribution is therefore quite broad and noisy.}
		\label{fig:ts}
\end{figure*}

In order to see how the dynamical time-step criterion works, we present the time-step distribution in four dark matter haloes, so called $\alpha\beta\gamma$-models \citep{1990ApJ...356..359H, 1993MNRAS.265..250D, 1996MNRAS.278..488Z}: with $\gamma = 1.5, 1.0, 0.5$ \& $0.0$ where $\gamma$ is the inner slope of the density profile $\rho(r) \propto r^{-\gamma}$. The outer slope is always $\beta = -3$. All haloes have a virial mass of $\Mvir = 10^{12}~h^{-1}{\Mo} = 1.429~\times10^{12}~\Mo$ ($h = 0.7$) and are represented by $\Nvir \approx 7.5\times10^6$ particles within the virial radius. This virial mass corresponds to a virial radius of $\rvir\approx289~\mathrm{kpc}$. We fix the concentration of the NFW profile \citep{1996ApJ...462..563N} to $c_{\mathrm{NFW}} = 10$ and the concentrations of the other profiles were chosen so that the maximum circular velocity is reached at the same radius in all haloes. The softening of the particles was $\epsilon = 0.1~\mathrm{kpc} \approx 3.5\times10^{-4}~\rvir$ in all haloes.

We compare our new time-step criterion based on the dynamical time with the standard criterion commonly used in $N$-body simulations. The standard criterion for selecting time-steps in $N$-body simulations is based on the acceleration of the particle. The rung, $n$, and time-step taken, $\Delta T_{\mathrm{S}}$, for the standard criterion is given by,
\begin{equation}\label{eq:TSstd}
\Delta T_{\mathrm{S}} = \frac{T_0}{2^n} \leq \eta_{\mathrm{S}} \sqrt{\frac{\epsilon}{|a(r)|}}~,
\end{equation}
where $\epsilon$ is the softening and $a$ the acceleration of the particle. By default, a value $\eta_{\mathrm{S}} = 0.2$ is generally used.
In spherically symmetric systems, the radius $r_{\mathrm{eq}}$ where the dynamical and standard criteria give the same time-step is given by
\begin{equation} \label{eq:req}
r_{\mathrm{eq}} = \frac{\eta_{\mathrm{S}}^2}{\eta_{\mathrm{D}}^2}~\epsilon~,
\end{equation}
independent of the form of the density profile.

In Fig. \ref{fig:ts} we plot the time-step criterion distribution of the particles for all haloes as a function of distance from the centre of the halo: between solid and long dashed lines the values for the dynamical time-step criterion (blue) and following the short dashed line the standard time-step criterion values (red). For a better overview, we only plot 0.1 per cent of the particles randomly selected from the total number of particles in each halo.

As we can see in Fig. \ref{fig:ts}, the standard criterion follows closely the theoretical curve (short dashed) with $|a(r)|$ calculated numerically. The dynamical time-step criterion also follows closely a band between $\eta_{\mathrm{D}} = 0.02 ... 0.03$ (long dashed and solid lines) for the theoretical curve with $\rho_{\mathrm{enc}} = \frac{M(r)}{r^3}$. The radius of equal time-steps with the parameters above results in $r_{\mathrm{eq}} = 4.444~\mathrm{kpc} \approx 1.5 \times 10^{-2}~\rvir$.

On the right side of each plot in Fig. \ref{fig:ts}, we also plot the time-step distribution in three bins at $r_i = \rvir, 10^{-1}~\rvir, 10^{-2}~\rvir$ of width 0.002 in logarithmic scale. For the dynamical time-stepping scheme, most of the particles lie in the band between the two curves. Of course, we expect a spread in our criterion since our add-up scheme does not recognize perfectly the geometry of the surrounding structure. We note however that the add-up scheme is generally more conservative for the choice of time-step at a given radius than the analytical value. In the two flat cases ($\gamma = 0.0$ and $\gamma = 0.5$), the bin at $10^{-2}~\rvir$ is close to the resolution limit of the halo and the distribution is therefore quite noisy and relatively broad. Most of the effect due to this broader distribution at small radii is absorbed by the block time-step scheme's way of discretizing the actual time-step taken (see also equation \ref{eq:TSdyn}); the time-step value provided by the criterion is just an upper limit.

With a rather conservative choice of $\eta_{\mathrm{D}} = 0.03$, we sample the orbit of a particle on a circular (tangential) orbit with at least $2 \pi / \eta_{\mathrm{D}} \approx 200$ steps. The second curve (short dashed) used a value of $\eta_{\mathrm{D}} = 0.02$ which corresponds to $2 \pi / \eta_{\mathrm{D}} \approx 300$ steps per circular orbit. 

The situation is a bit more complicated for particles on perfect radial orbits. For a homogeneous sphere, the particle will oscillate through the centre of the sphere and describe therefore a harmonic oscillator with period $T = \sqrt{3 \pi/(G \rho)}$ where $\rho$ is the homogeneous density of the sphere.\footnote{Note that \cite{1987gady.book.....B} defines the dynamical time as the time needed by the particle to reach the centre which means $T_{\mathrm{dyn}} \equiv \frac{1}{4}T =\sqrt{3 \pi/(16 G \rho)}$.} If we take this value with $\rho_{\mathrm{enc}} = \rho$ we get for a complete radial orbit between $\sqrt{3 \pi}/0.03 \approx 100$ and $\sqrt{3 \pi}/0.02 \approx 150$ steps per oscillation. Since our time-step criterion is very adaptive with radius, the dynamical time will decrease in a steep density profile when the particle approaches the centre, so that the effective number of steps is even higher.

The main disadvantage of the standard time-step criterion (\ref{eq:TSstd}) is the bad adaptivity with radius, i.e. the particles are distributed over relatively few rungs. Especially in the NFW profile, the particles inside about ten per cent of the virial radius are all on the same time-step. For flatter central profiles with $\gamma<1$ where $\rho(r) \propto r^{-\gamma}$, the time-steps even increase inside the radius where the acceleration has its maximum, in clear contradiction to the behaviour of the dynamical time!
The asymptotic radial behaviour in the central region of the standard time-stepping criterion is given by
\begin{equation} \label{eq:tstdbeh}
\Delta T_{\mathrm{S}} \propto \sqrt{\frac{1}{|a(r)|}} \propto r^{\frac{\gamma-1}{2}} \quad (\gamma < 3)~,
\end{equation}
where $\gamma$ is the inner slope of the density profile. In contrast, the dynamical time-stepping scheme has the following dependence
\begin{equation}\label{eq:tdynbeh}
\Delta T_{\mathrm{D}} \propto \sqrt{\frac{r^3}{M(r)}} \propto r^{\frac{\gamma}{2}} \quad (\gamma < 3)~.
\end{equation}
The standard criterion can only obtain the correct choice of time-step in the central regions by shifting the above relation, either by choosing a small softening for these particles, or by reducing $\eta_{\mathrm{S}}$. This automatically leads to an immense computational expense due to the overly conservative time-steps in the outer parts of the halo or even wrong physical behaviour due to the choice of too small softening for the physical problem (e.g. undesired scattering of particles). The radial scaling of the standard criterion makes it ill suited to the study of the centre of galaxies and in other situations where a very large dynamic range in density needs to be evolved correctly. The dynamical time-stepping technique we present is a much more universal approach to choosing time-steps in self-gravitating systems, since the basic parameters of the method, such as the angle for adding up mass contributions, once determined, are kept fixed for different simulations.

\subsection{Elliptic 2-body orbits}

\begin{figure*}
	\centering
		\includegraphics[width=0.495\textwidth]{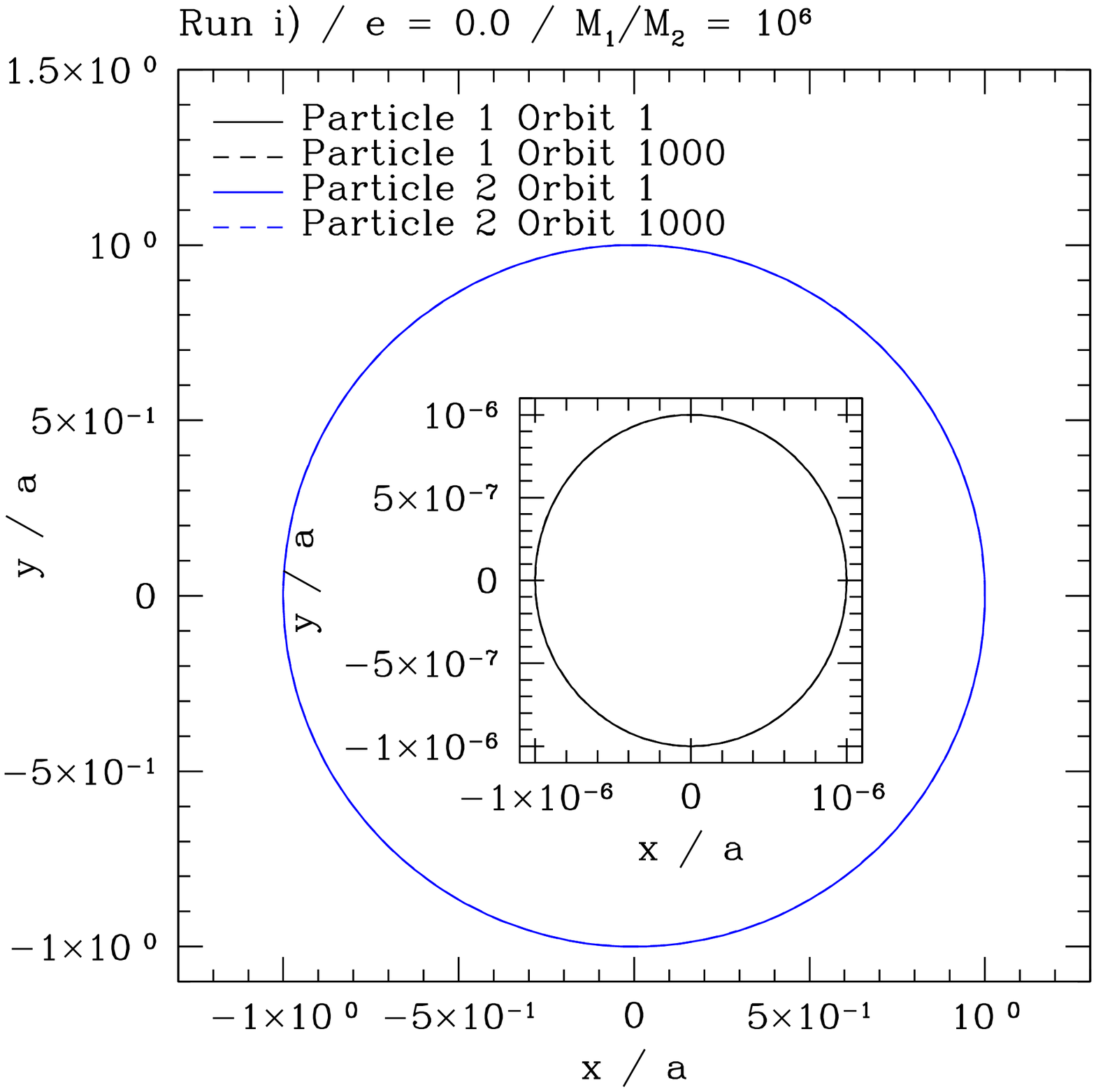} \hfill
		\includegraphics[width=0.495\textwidth]{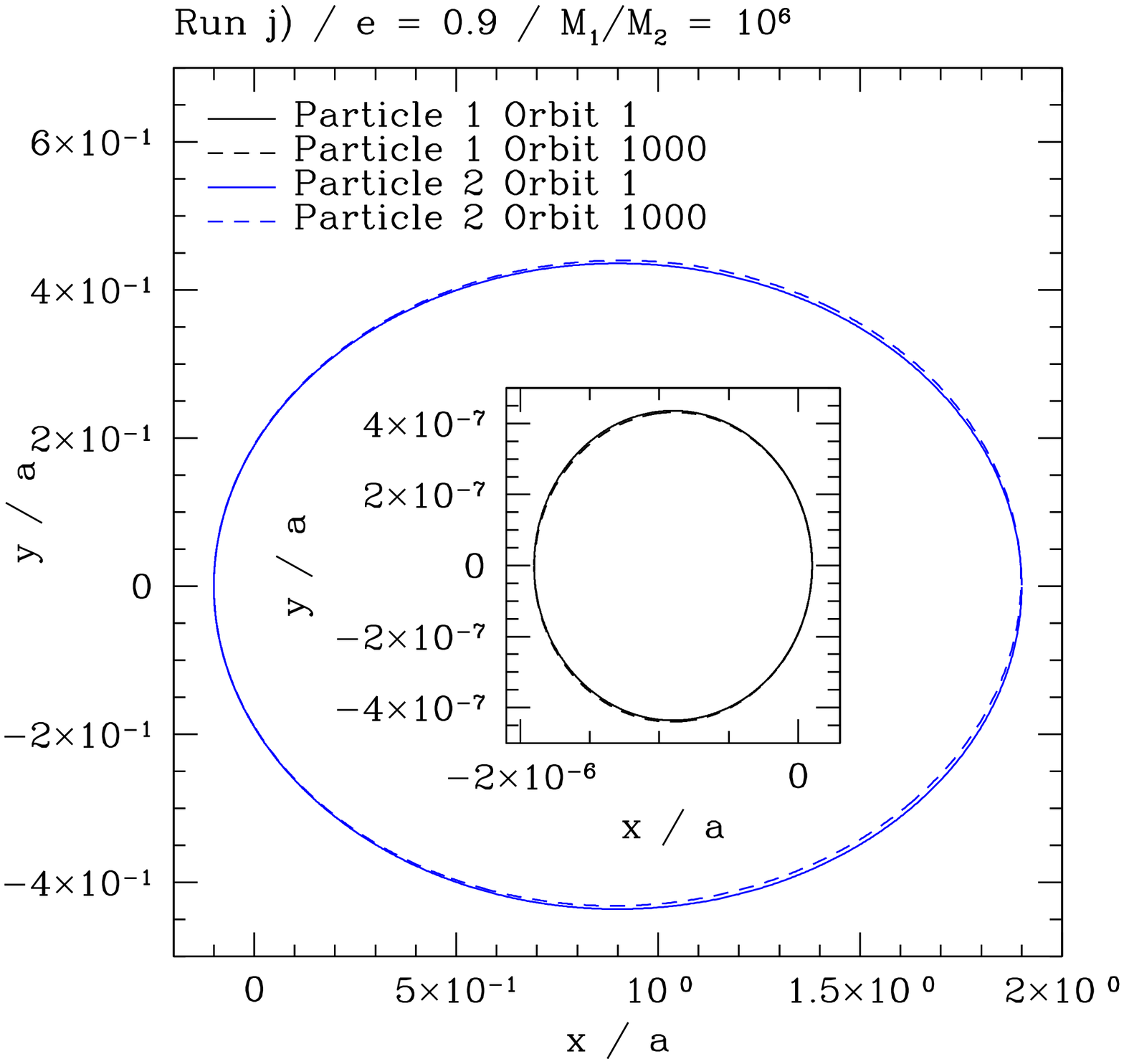} \\
		\includegraphics[width=0.495\textwidth]{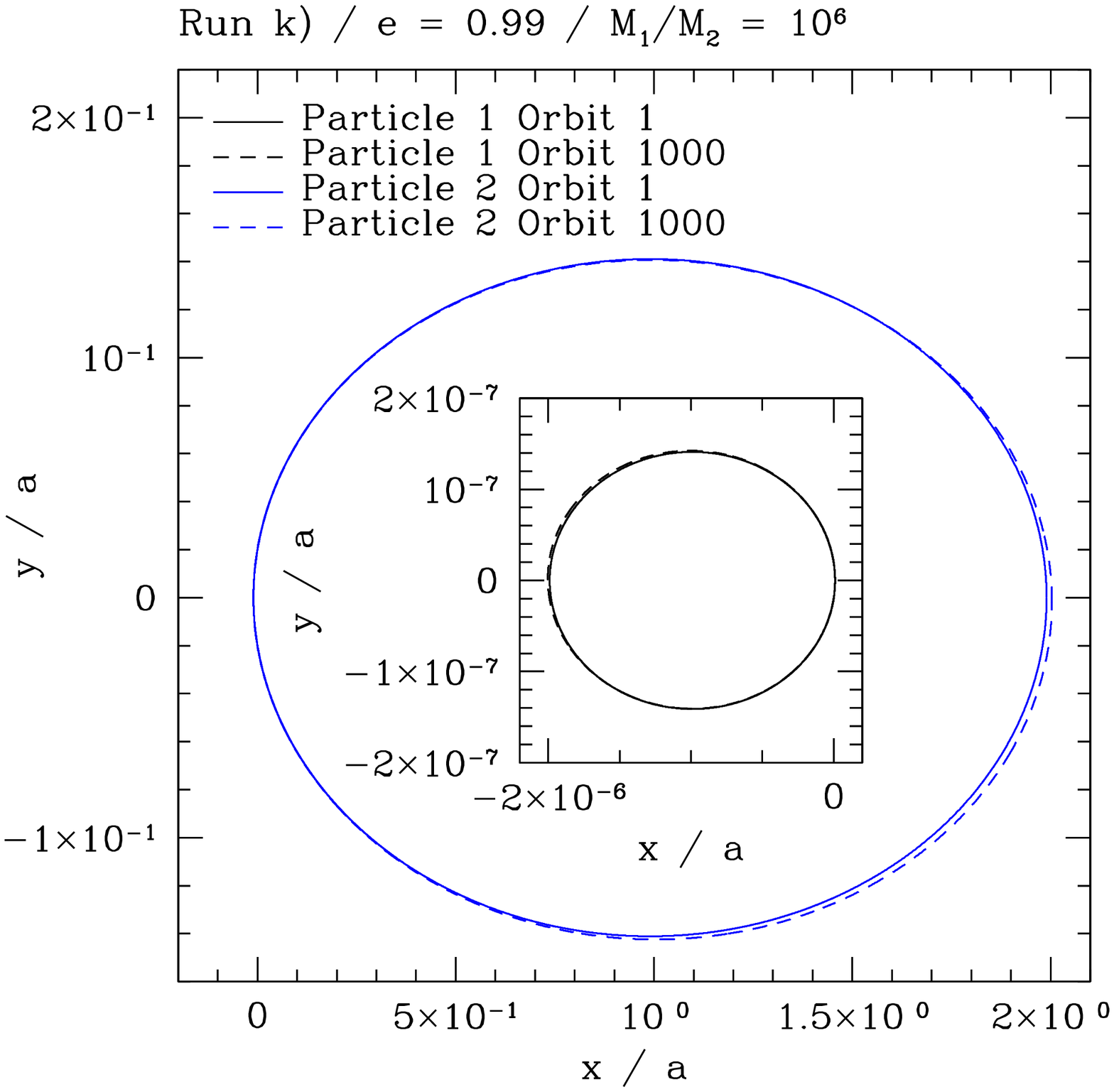} \hfill
		\includegraphics[width=0.495\textwidth]{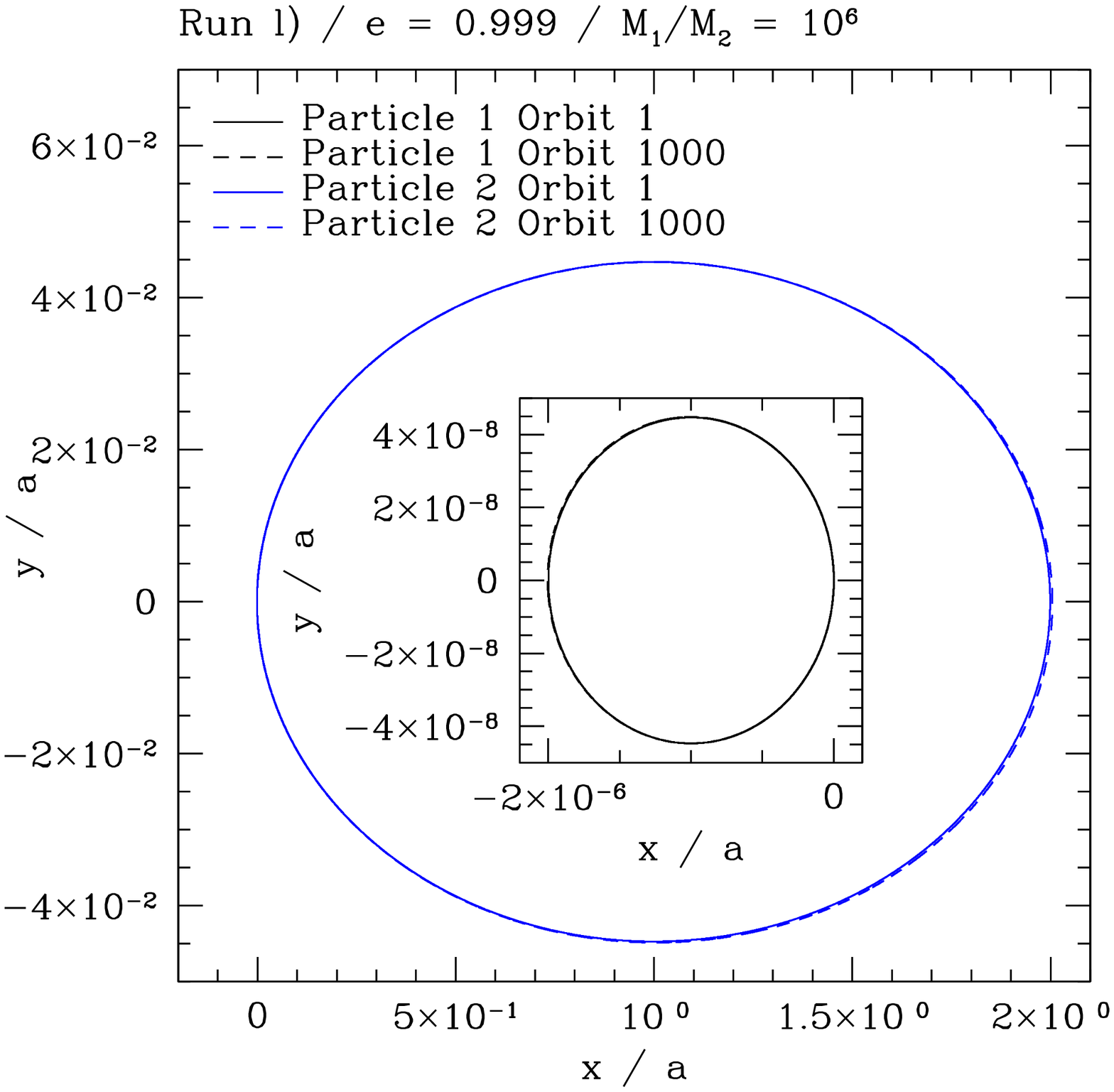}
		\caption{Top left: run i), top right: run j), bottom left: run k), bottom right: run l) from Table \ref{tab:2bodyell}. We plot the orbit from $0$ to $T_{\mathrm{K}}$ (orbit 1) and the orbit from $999~T_{\mathrm{K}}$ to $1000~T_{\mathrm{K}}$ (orbit 1000) for each run. All runs used the dynamical time-stepping scheme with eccentricity correction. We deliberately did not plot the scaling of the two axes constrained in order to make the small deviations visible.}
		\label{fig:2bodyell}
\end{figure*}

In order to quantify the performance of our adaptive dynamical time-stepping criterion in the scattering regime, we performed a series of simulations studying the behaviour of high eccentricity 2-body Kepler orbits. After choosing the masses $M_1$ and $M_2$ of the two bodies, all other quantities are fixed, i.e. the orbital time of the Kepler orbit is given by,
\begin{equation}
T_{\mathrm{K}}\equiv\sqrt{\frac{a^3 (2 \pi)^2 \mu}{G M_1 M_2}} = 2 \pi \sqrt{\frac{a^3}{G (M_1 + M_2)}}
\end{equation}
and the initial total energy is calculated by,
\begin{equation}
E_0 \equiv -\frac{G M_1 M_2}{2a}~.
\end{equation}
We chose a unit system where Newton's gravitational constant $G \equiv 1$ and we fix the orbit in the same way, i.e. the semi-major axis is always $a \equiv 1$. The softening $\epsilon$ of the two particles was set to $0.1~d_{\mathrm{peri}}$ in all cases where $d_{\mathrm{peri}} \equiv a (1-e)$ is the periapsis distance of the Kepler orbit. \textsc{Pkdgrav} treats the forces completely Newtonian if the two particles have a distance larger than $2\epsilon$ which is therefore always the case in these test runs. Initially, the particles were set in a coordinate system where the centre of mass is at rest at the origin and the two particles were at apoapsis configuration along the $x$-axis. A short summary of the parameters can be found in Table \ref{tab:2bodyell}.

\begin{table*}
	\caption{Summary of elliptic 2-body orbit runs a) - p).}
	\label{tab:2bodyell}
	\begin{tabular}{ccccccccc}
		\hline
		Run &$M_1/M_2$& $e$   &$T_{\mathrm{K}}$      &$E_0$            & $(E_{1000}-E_0)/|E_0|$ & $N$ & time-stepping scheme & compare with run \\
		\hline
		a)  & 1       & 0     & 4.443                &$-5\times10^{-1}$& $2.238\times10^{-9}$   & 256 & D / ec & \\
		b)  & 1       & 0.9   & 4.443                &$-5\times10^{-1}$& $-7.278\times10^{-4}$  & 2110 & D / ec & m), n), o)\\
		c)  & 1       & 0.99  & 4.443                &$-5\times10^{-1}$& $6.648\times10^{-3}$   & 9394 & D / ec & \\
		d)  & 1       & 0.999 & 4.443                &$-5\times10^{-1}$& $2.804\times10^{-3}$   & 39143 & D / ec & \\
	  e)  & $10^3$  & 0     & $1.985\times10^{-1}$ &$-5\times10^{2}$ & $2.254\times10^{-9}$   & 256 & D / ec & \\
		f)  & $10^3$  & 0.9   & $1.985\times10^{-1}$ &$-5\times10^{2}$ & $-7.014\times10^{-4}$  & 2110 & D / ec & \\
		g)  & $10^3$  & 0.99  & $1.985\times10^{-1}$ &$-5\times10^{2}$ & $6.648\times10^{-3}$   & 9394 & D / ec & \\
		h)  & $10^3$  & 0.999 & $1.985\times10^{-1}$ &$-5\times10^{2}$ & $2.804\times10^{-3}$   & 39142 & D / ec & \\
		i)  & $10^6$  & 0     & $6.283\times10^{-3}$ &$-5\times10^{5}$ & $2.242\times10^{-9}$   & 256 & D / ec & \\
		j)  & $10^6$  & 0.9   & $6.283\times10^{-3}$ &$-5\times10^{5}$ & $-7.218\times10^{-4}$  & 2110 & D / ec & p) \\
		k)  & $10^6$  & 0.99  & $6.283\times10^{-3}$ &$-5\times10^{5}$ & $6.653\times10^{-3}$   & 9394 & D / ec & \\
		l)  & $10^6$  & 0.999 & $6.283\times10^{-3}$ &$-5\times10^{5}$ & $2.802\times10^{-3}$   & 39142 & D / ec & \\
		m)  & 1       & 0.9   & 4.443                &$-5\times10^{-1}$& $1.730\times10^{-1}$   & 398 & D / nec & b) \\
		n)  & 1       & 0.9   & 4.443                &$-5\times10^{-1}$& $-1.263$               & 316 & S  & b) \\
		o)  & 1       & 0.9   & 4.443                &$-5\times10^{-1}$& $1.895\times10^{-3}$   & 2107 & S / $\eta_{\mathrm{S}} = 0.029$ & b) \\
		p)	& $10^6$  & 0.9   & $6.283\times10^{-3}$ &$-5\times10^{5}$ & $1.913$                & 3197 & S & j) \\
		\hline
	\end{tabular}
	\medskip
	\begin{flushleft}
	The columns are mass ratio $M_1/M_2$, eccentricity $e$, orbital time $T_{\mathrm{K}}$, initial energy $E_0$, relative energy change $(E_{1000}-E_0)/|E_0|$, number of steps during the first orbit $N$, time-stepping scheme: D - dynamical, S - standard, ec - eccentricity correction, nec - no eccentricity correction, run to compare with: b) with m), n) and o) and j) with p).
	\end{flushleft}
\end{table*}

We let each run evolve for $1000~T_{\mathrm{K}}$ ($T_{\mathrm{K}}$ was also the basic or longest time-step in the block time-stepping scheme, $T_0$), measured the total energy $E_{1000}$ after the end of the run and calculated the relative energy shift $(E_{1000}-E_0)/|E_0|$. These values are also listed in Table \ref{tab:2bodyell}. From Table \ref{tab:2bodyell}, we see that even for a high eccentricity ($e = 0.999$) orbit we still have relative energy conservation on the level of $\approx10^{-6}$ per orbit. The general behaviour can also be seen in Fig. \ref{fig:2bodyell} where we plot the orbit from $0$ to $T_{\mathrm{K}}$ and the orbit from $999~T_{\mathrm{K}}$ to $1000~T_{\mathrm{K}}$ for each of the different runs i) - l). The two orbits lie nearly on top of each other except for the $e=0.99$ case, where the energy gain in the integration was the largest, we can see a small deviation. When we compare the runs with different mass ratios, we see that the relative energy conservation is nearly the same, i.e. the relative energy conservation depends only on the geometry of the orbit.

\begin{figure*}
	\centering
		\includegraphics[width=0.495\textwidth]{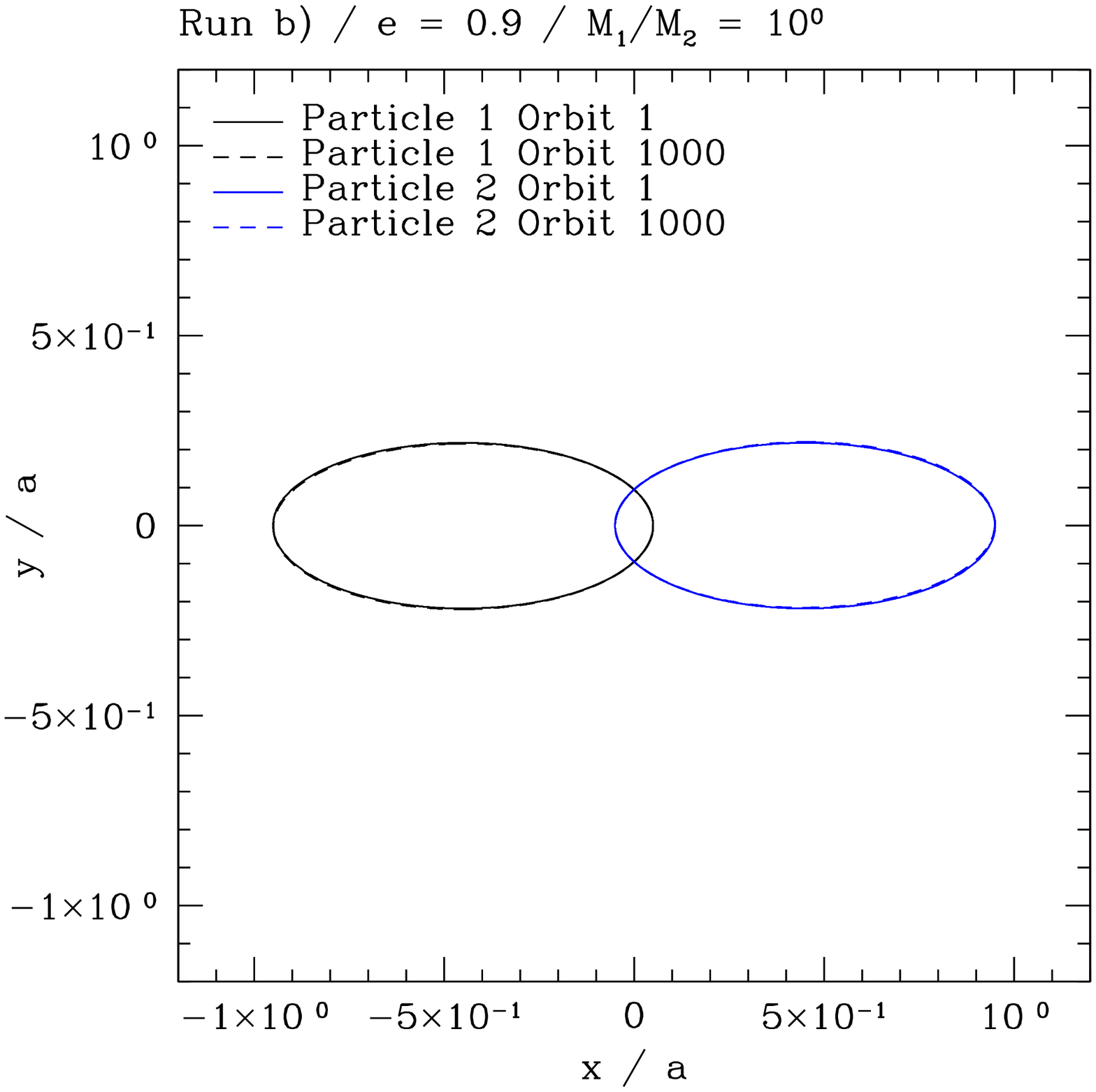} \hfill
		\includegraphics[width=0.495\textwidth]{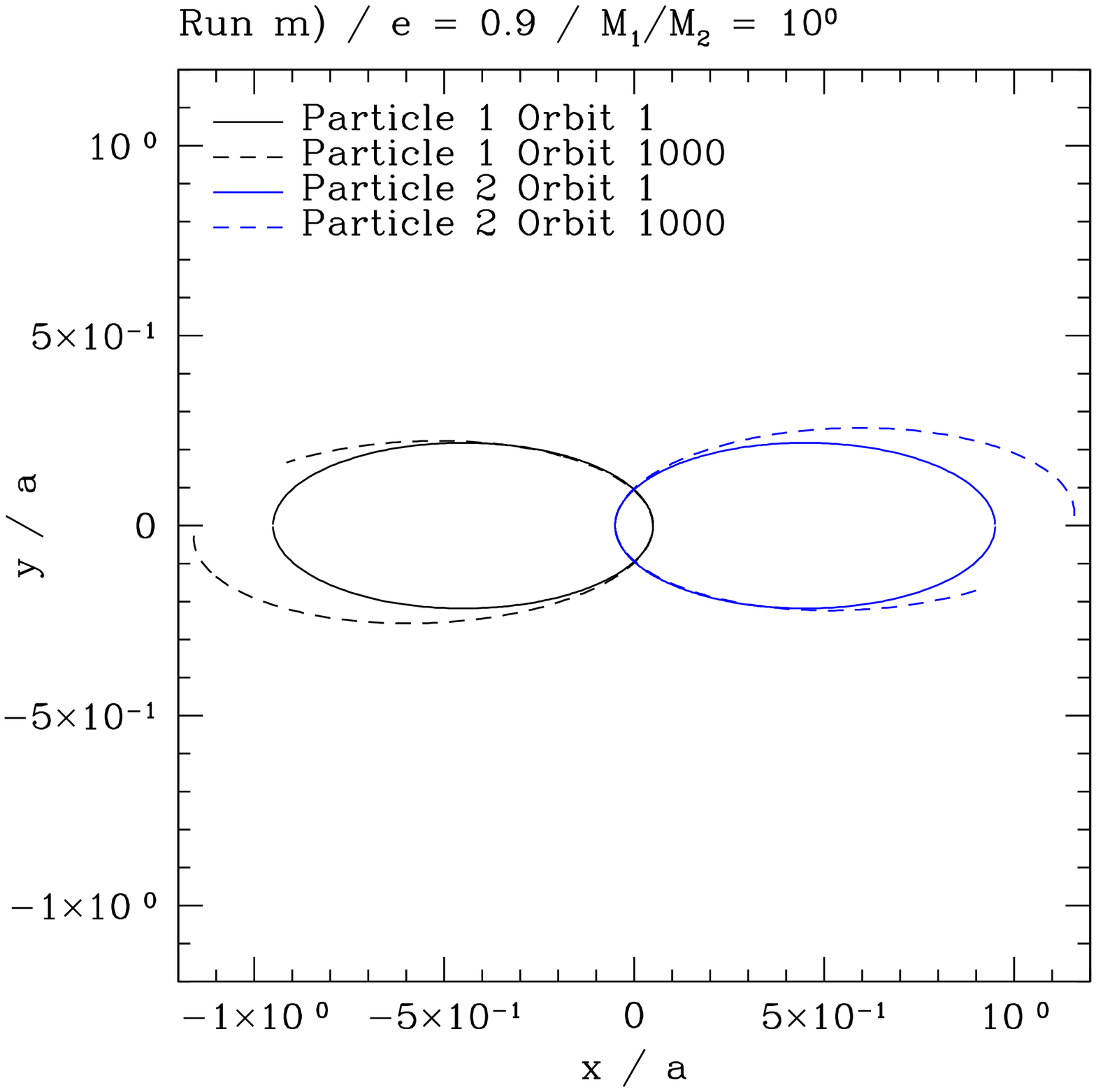} \\
		\includegraphics[width=0.495\textwidth]{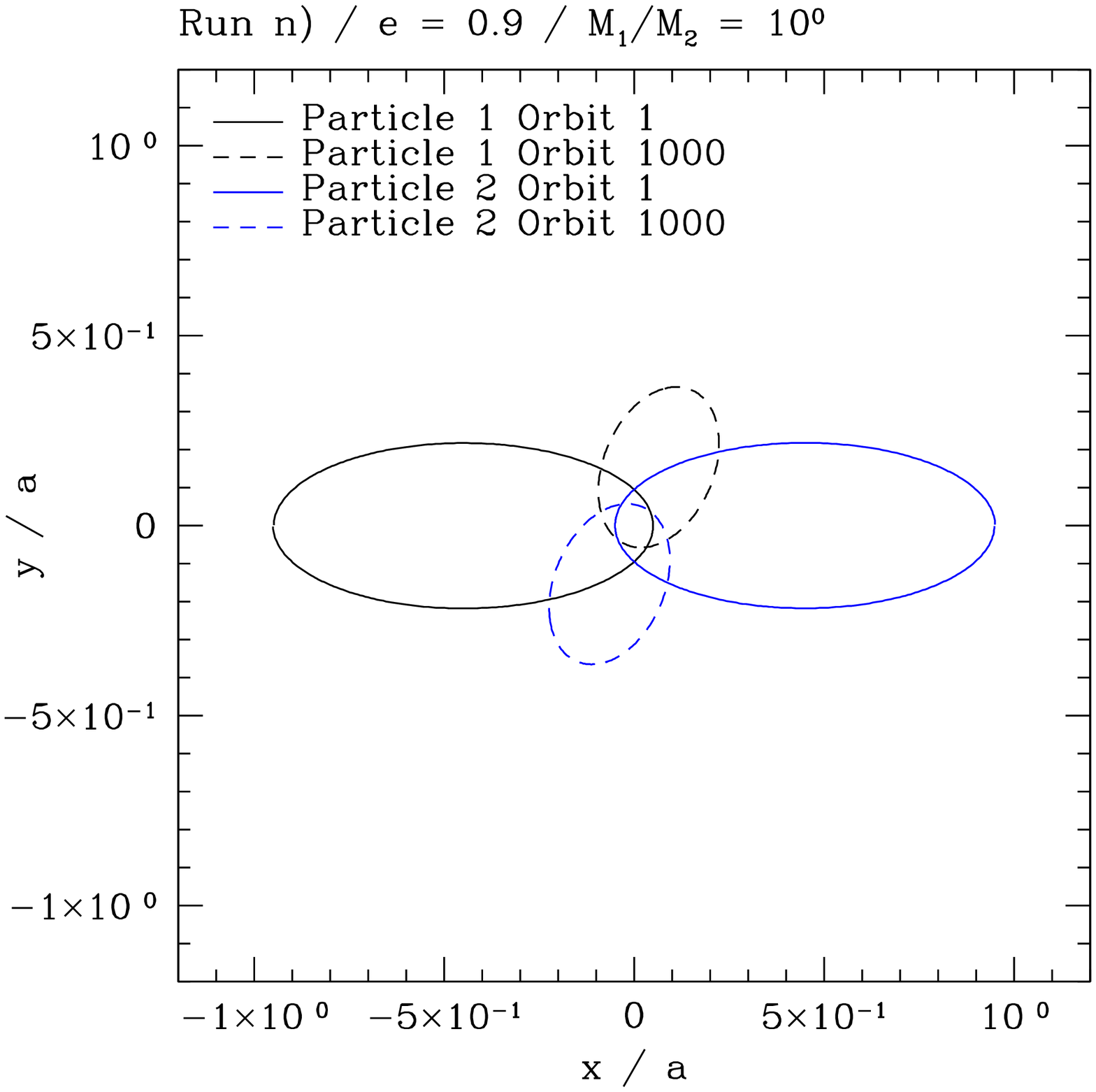} \hfill
		\includegraphics[width=0.495\textwidth]{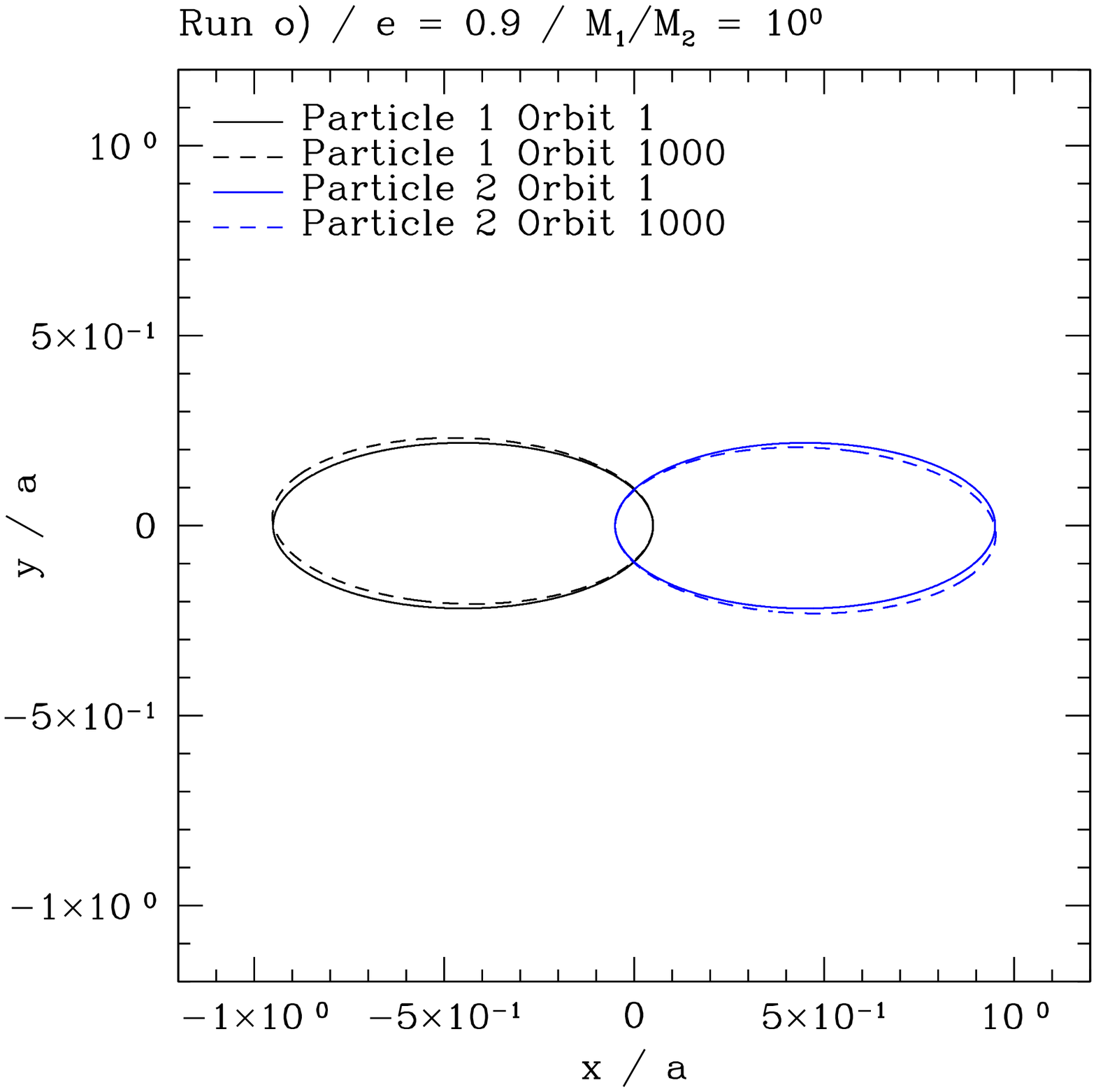} \\
		\caption{Top left: run b), top right: run m), bottom left: run n), bottom right: run o) from Table \ref{tab:2bodyell}. Run b) shows the dynamical time-stepping case with eccentricity correction. The orbit is perfectly followed. In run m), we can nicely see the energy gain visually due to the lack of the eccentricity correction in the time-step criterion. Run n) where we tried to resolve a $e=0.9$ orbit analogue to run b) with the standard time-step criterion. As can be seen, the orbit becomes more circular and the orbital plane starts to rotate which is completely unphysical. In run o), we see a run where we used a smaller value of $\eta_{\mathrm{S}} = 0.029$ so that the standard criterion initially makes an equal number of steps per orbit as the dynamical time-stepping scheme in run b). This helps a lot but it does still not perform as good as the dynamical time-stepping scheme.}
		\label{fig:2bodyell2}
\end{figure*}

In order to illustrate the robustness of our method we have performed some further tests. In run m), we switched off the eccentricity correction in the symmetrised dynamical time-stepping (\ref{eq:REPsym}), i.e. $C(e) =1$. We see that the energy gain over 1000 $T_{\mathrm{K}}$ is $\approx 240$ times larger than in the corresponding run b) with eccentricity correction. This can also be seen visually in Fig. \ref{fig:2bodyell2}. Due to the energy gain, the orbits of both particles become wider. 

In run n) we tried to resolve the orbit with the standard time-stepping criterion given by equation (\ref{eq:TSstd}). This criterion depends on the softening length $\epsilon$ of the particle and is therefore certainly not ideal in the gravitational scattering regime. We've chosen the same value as in run b) where we had $\epsilon = 0.1~d_{\mathrm{peri}} = 0.01$. From Fig. \ref{fig:2bodyell2} we see immediately when comparing run n) and b) that the standard criterion can not capture the dynamics of the orbit. The orbits of the two particles become more circular and we get a rotation of the whole system. If we had chosen a somewhat larger softening, so that it is still smaller than half the periapsis distance, it would even look worse since the standard time-stepping scheme directly depends on the softening length $\epsilon$ while the dynamical time-stepping scheme would still perform equally well since it has no such dependence.

Of course, the dynamical time-stepping criterion with eccentricity correction uses a lot more steps per orbit than the standard criterion with $\eta_{\mathrm{S}} = 0.2$. Therefore we tried in run o) a run with an equal number of steps per orbit as run b). This is reached for a value $\eta = 0.029$. The energy conservation is still not as good as in the case of the dynamical time-stepping with eccentricity correction b) and there is a small amount of precession of the periapsis.

\begin{figure}
	\centering
		\includegraphics[width=\columnwidth]{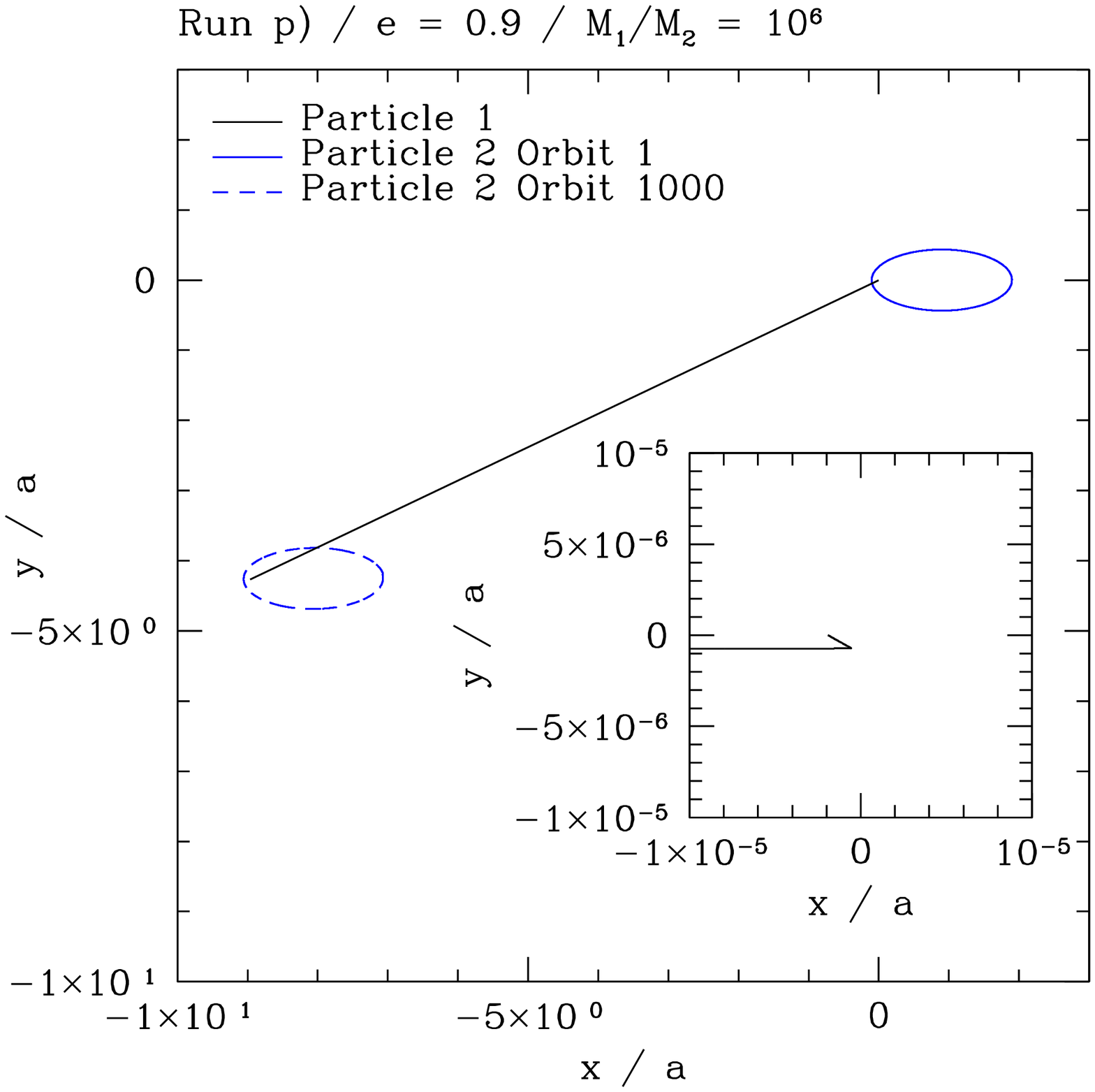}
		\caption{Run p) where we tried to resolve an $e=0.9$, high mass ratio orbit with the standard time-stepping criterion. The heavy particle gets an immense kick and the total energy becomes positive. Thus, the heavy particle drags the light particle with it and the whole system drifts apart. The standard criterion therefore fails completely to follow the orbit correctly.}
	\label{fig:2be09mr6std.orbit}
\end{figure}

The whole situation becomes even worse when we try to resolve 2-body orbits with unequal mass particles. Since the standard time-stepping is not symmetric due to the asymmetry in acceleration, it is not able to resolve a high mass ratio 2-body orbit correctly and fails completely. This is shown in Fig. \ref{fig:2be09mr6std.orbit}. Although the light particle makes $N = 3197$ steps in the first orbit (here $N$ in run p) denotes only the number of steps of the light particle in Table \ref{tab:2bodyell}), the heavy particle takes a much larger first step than the light particle. Of course when then the light particle approaches, the heavy particle gets an immense kick, the total energy becomes positive and the whole system drifts apart. 

\subsection{Hyperbolic 2-body orbits}

In a similar way, we also tested the new dynamical time-stepping scheme for hyperbolic orbits. Initial conditions were set up such that the line connecting the two particles encloses an angle of $\frac{\pi}{6}$ with the semi-major axis (symmetry axis) of the hyperbola. The time for the particle to reach the periapsis of the orbit is given by
\begin{equation}
T_{\mathrm{H}} = \int_{\frac{\pi}{6}}^\pi \frac{\mu ~ r^2(\phi)}{L} d\phi
\end{equation}
where $r(\phi)$ describes the angle dependent relative separation of the two bodies. The initial conditions used the same unit system as the elliptic orbit tests and we again set the softening to $0.1~d_{\mathrm{peri}}$ where $d_{\mathrm{peri}} \equiv a (e-1)$ for hyperbolic orbits. A summary of the different runs can be found in Table \ref{tab:2bodyhyp}.

\begin{table*}
	\caption{Summary of hyperbolic 2-body orbit runs A) - K).}
	\label{tab:2bodyhyp}
	\begin{tabular}{ccccccccccc}
		\hline
		Run &$M_1/M_2$& $e$   & $T_{\mathrm{H}}$     &$E_0$           & $(E_{1000}-E_0)/|E_0|$ &$N$ & time-stepping scheme & compare with run\\
		\hline
		A)  & 1       & 1.1   & $2.150$              &$5\times10^{-1}$& $3.009\times10^{-3}$ & 1713 & D / ec & \\
		B)  & 1       & 1.01  & $2.274\times10^{-2}$ &$5\times10^{-1}$& $2.119\times10^{-3}$ & 4935 & D / ec & \\
		C)  & 1       & 1.001 & $6.709\times10^{-4}$ &$5\times10^{-1}$& $-1.298\times10^{-2}$& 15177 &D / ec & \\
		D)  & $10^3$  & 1.1   & $9.610\times10^{-2}$ &$5\times10^{2}$ & $3.009\times10^{-3}$ & 1713 & D / ec & \\
		E)	& $10^3$  & 1.01  & $1.017\times10^{-3}$ &$5\times10^{2}$ & $2.119\times10^{-3}$ & 4935 & D / ec & \\
		F)	& $10^3$  & 1.001 & $2.999\times10^{-5}$ &$5\times10^{2}$ & $-1.298\times10^{-2}$& 15178 & D / ec & \\
		G)	& $10^6$  & 1.1   & $3.041\times10^{-3}$ &$5\times10^{5}$ & $3.009\times10^{-3}$ & 1713 & D / ec & J), K)\\
		H)	& $10^6$  & 1.01  & $3.216\times10^{-5}$ &$5\times10^{5}$ & $2.118\times10^{-3}$ & 4935 & D / ec & \\
		I)	& $10^6$  & 1.001 & $9.488\times10^{-7}$ &$5\times10^{5}$ & $-1.298\times10^{-2}$& 15177 & D / ec & \\
		J)	& $10^6$  & 1.1   & $3.041\times10^{-3}$ &$5\times10^{5}$ & $8.336\times10^{-1}$ & 296 & D / nec & G) \\       
		K)	& $10^6$  & 1.1   & $3.041\times10^{-3}$ &$5\times10^{5}$ & $1.360\times10^{-1}$ & 333 & S & G) \\
		\hline
	\end{tabular}
	\medskip
	\begin{flushleft}
	The columns are mass ratio $M_1/M_2$, eccentricity $e$, time to reach the pericentre $T_{\mathrm{H}}$, initial energy $E_0$, relative energy change $(E_{1000}-E_0)/|E_0|$, number of steps during the first orbit $N$, time-stepping scheme: D - dynamical, S - standard, ec - eccentricity correction, nec - no eccentricity correction, run to compare with: G) with J) and K).
	\end{flushleft}
\end{table*}

In order to get an integrated effect, we mirrored the velocities of the particles after $2~T_{\mathrm{H}}$ and let the runs evolve in total for $2000~T_{\mathrm{H}}$ in order to get 1000 pericentre passages.

\begin{figure}
	\centering
		\includegraphics[width=\columnwidth]{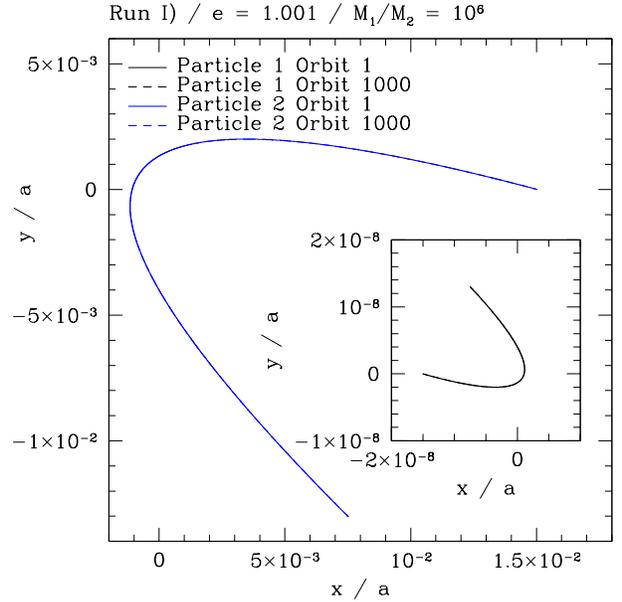}
		\caption{The most extreme case, run I), with mass ratio of $10^6$ and eccentricity $e=1.001$. The dynamical time-stepping scheme with eccentricity correction is used. There is no significant evolution over 1000 repeated pericentre passages.}
		\label{fig:2bodyhyp}
\end{figure}

In Fig. \ref{fig:2bodyhyp}, we plot the most extreme case, run I), with mass ratio of $10^6$ and eccentricity $e=1.001$. Over 1000 pericentre passages, there is no visible evolution of the orbit and the relative energy change is of order $O(10^{-5})$ per orbit. For the other cases with lower eccentricities and mass ratios, the new dynamical time-stepping scheme works optimally and we do not show the other orbits here.

\begin{figure*}
	\centering
		\includegraphics[width=0.495\textwidth]{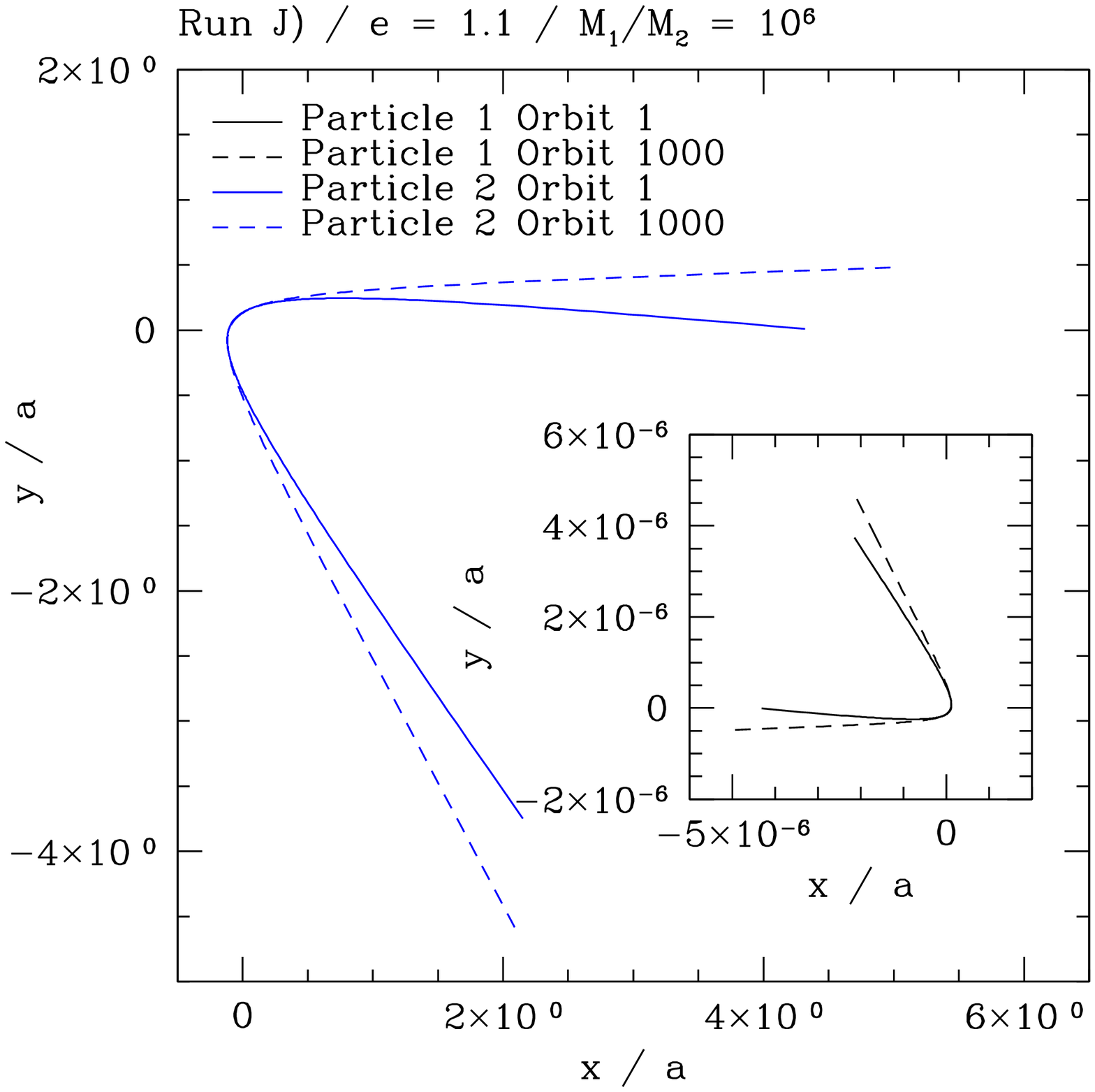} \hfill
		\includegraphics[width=0.495\textwidth]{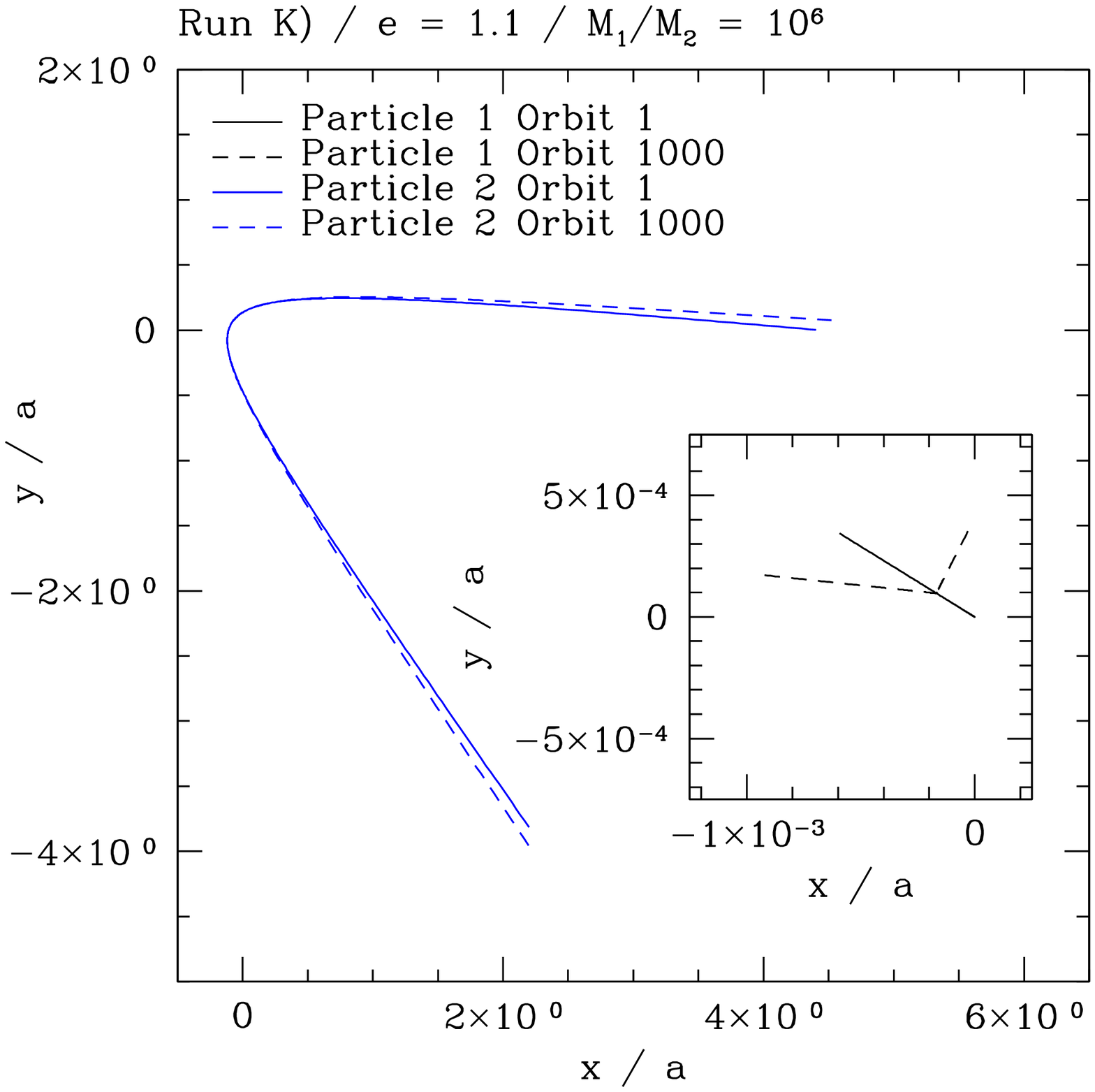} \\
		\caption{Left: run J), right: run K) from Table \ref{tab:2bodyhyp}. In run J) we see the energy gain if the orbit is not followed correctly by not using the eccentricity correction in the dynamical time-stepping scheme. On the right, we see run K) where we used the standard time-stepping scheme. The heavy particle is wandering around in a much larger area than allowed, as can be seen by comparing the scales of the two inset plots.}
		\label{fig:2bodyhyp2}
\end{figure*}

We tried again to resolve high mass ratio orbits without eccentricity correction (run J) and the standard time-stepping scheme (run K). The results can be seen in Fig. \ref{fig:2bodyhyp2}. If we do not correct for the eccentricity (as in run J), the particles gain energy and the orbits become wider. In run K) we see the behaviour of the standard time-stepping scheme. Due to the large mass ratio, the acceleration of the light particle is quite large and therefore it follows a qualitatively correct orbit due to the small time-steps. This is similar to the case p) of the elliptic 2-body orbits where the light particle describes an elliptical orbit about the massive particle, even though this massive particle gets a spurious kick. Once again, the massive particle has a completely incorrect oribit wandering around in a very large area due to the spurious kicks (compare the scales of the inset plots in runs J and K).
The lack of momentum conservation in such cases results in this contrasting behaviour of the two different mass particles. The symmetrization of the time-step criterion restores momentum conservation between the two particles involved in the gravitational scattering event.

\subsection{Cosmological structure formation}

We also tested the performance of the dynamical time-stepping scheme in a cosmological structure formation run. For this purpose, we used a fiducial simulation of the Virgo cluster \citep{1998MNRAS.300..146G} in a cosmological framework with $\Omega_{\mathrm{M}} = 1$, no cosmological constant i.e. $\Omega_{\Lambda} = 0$ and $H_0$ = 50 km s$^{-1}$ Mpc$^{-1}$. The simulation cube had a box length of $L$ = 100 Mpc and the total mass in the cube was $M_{\mathrm{tot}} = 6.937\times10^{16}~\Mo$. The cluster was resolved using the standard refinement technique \citep{1993ApJ...412..455K, 1994MNRAS.270L..71K} so that the particle mass in the highest resolution region was $8.604\times10^8~\Mo$ and the softening length of the lightest particles was $\epsilon = 5~\mathrm{kpc}$. The total number of particles was $1.314\times10^6$. The simulation started at redshift $z=69$ and we evolved the cluster to redshift $z=0$ with three different time-stepping schemes: one dynamical and one standard time-stepping run and, for comparison, a fixed time-step scheme with 300000 time-steps from $z=69$ to $z=0$ (this corresponds to 20000 time-steps down to redshift $z=5$ in the above described cosmology). With this choice, the fixed time-step length corresponds approximately to the smallest time-step chosen by the dynamical criterion during the whole run. Only for a few particles, the dynamical scheme did choose smaller steps than this fixed time-stepping run. 

\begin{figure}
	\centering
		\includegraphics[width=\columnwidth]{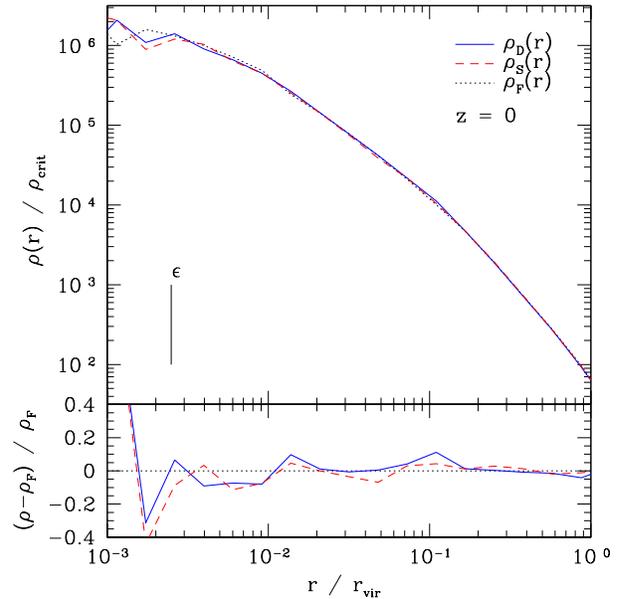}
	\caption{Density profiles of the three runs of the Virgo cluster with the different time-stepping schemes at final redshift $z=0$: $\rho_{\mathrm{D}}(r)$ is the radial density of the run with the dynamical time-stepping scheme, $\rho_{\mathrm{S}}(r)$ the density profile of the run with the standard time-stepping scheme and $\rho_{\mathrm{F}}(r)$ the profile of the fixed time-step run. The profiles are normalised with respect to the critical density $\rho_{\mathrm{crit}}$. On the top panel, the absolute values and on the lower panel the relative differences $(\rho(r) - \rho_{\mathrm{F}}(r))/\rho_{\mathrm{F}}(r)$ are plotted.}
	\label{fig:virgo.pro.z0}
\end{figure}

The virial radius of the resulting cluster was in all cases $\rvir \approx$ 2 Mpc (overdensity $\approx$ 200) and had a final mass of $M_{\mathrm{Cluster}} \approx 4.3\times 10^{14}~\Mo$. In Fig. \ref{fig:virgo.pro.z0}, we plot on the top panel the radial density profile for the three runs at redshift $z=0$. Here $\rho_{\mathrm{D}}(r)$ is the radial density of the run with the dynamical time-stepping scheme, $\rho_{\mathrm{S}}(r)$ the density profile of the run with the standard time-stepping scheme and $\rho_{\mathrm{F}}(r)$ the profile of the fixed time-step run. In the lower panel, the relative difference $(\rho(r) - \rho_{\mathrm{F}}(r))/\rho_{\mathrm{F}}(r)$ is also plotted. The softening of the highest resolution particles correspond to $\epsilon = 5~\mathrm{kpc} \approx 2.5 \times 10^{-3}~\rvir$. As we can see in Fig. \ref{fig:virgo.pro.z0}, the same radial density profile is obtained for the final cluster in this cosmological simulation.

\begin{figure*}
	\centering
		\includegraphics[width=0.495\textwidth]{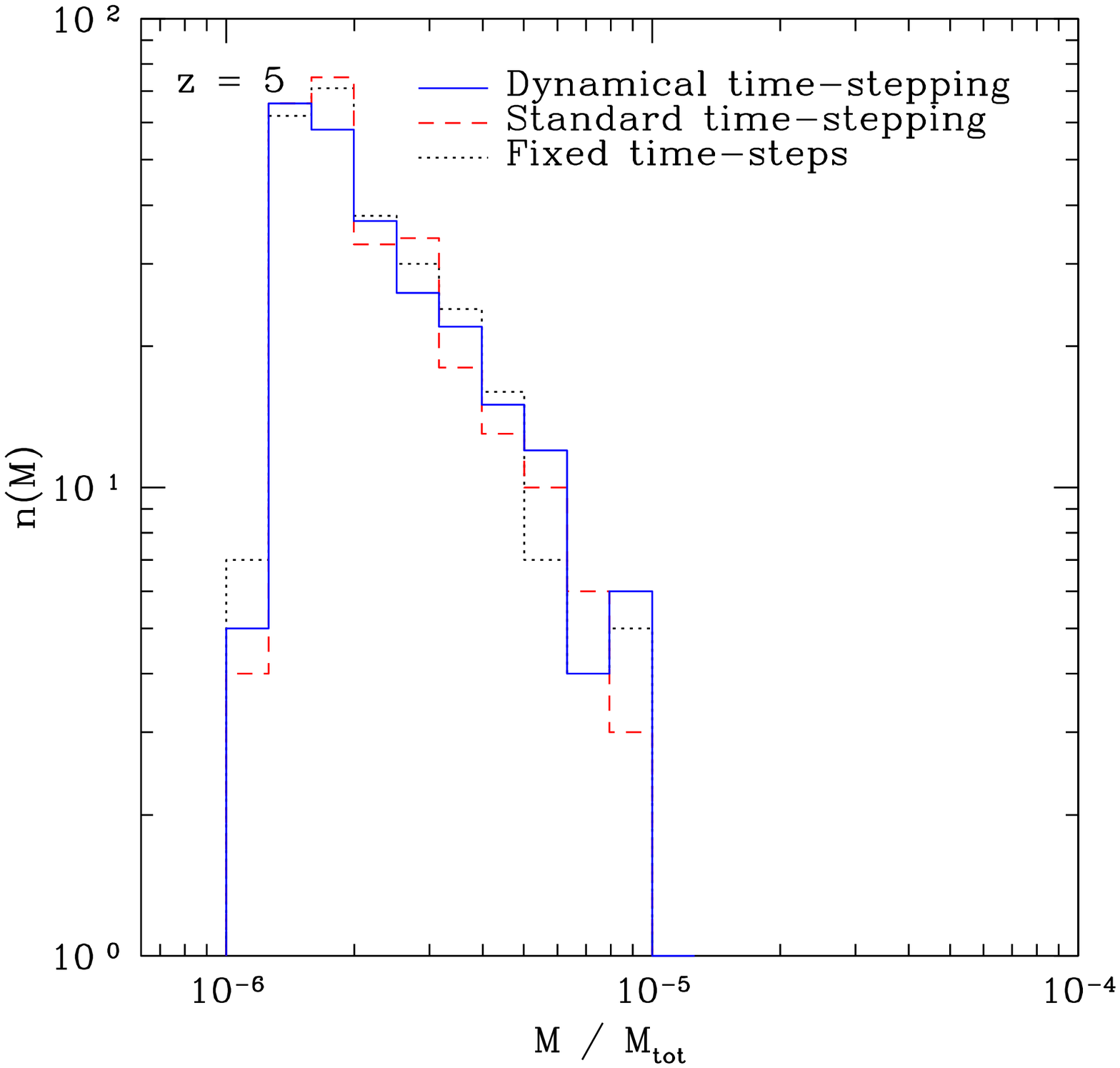} \hfill
		\includegraphics[width=0.495\textwidth]{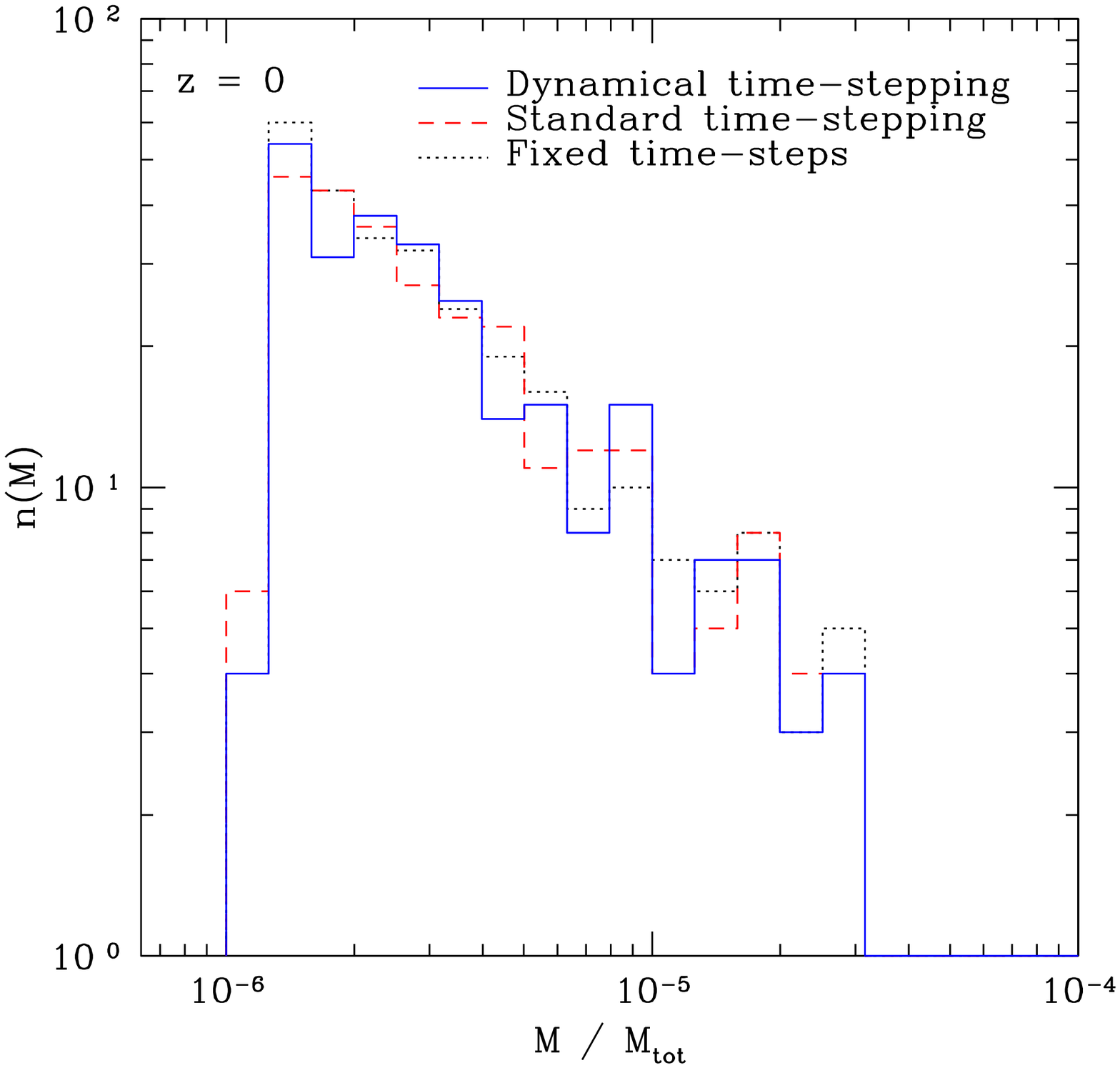}
		\caption{Substructure mass function of the Virgo cluster run at redshift $z=5$ (left) and $z=0$ (right). There is no substantial difference between the runs with different time-stepping schemes visible.}
	\label{fig:virgo}
\end{figure*}

We also compared the substructure mass function at redshifts $z=0$ and $z=5$ for the three runs. For that, we used the group finding software skid\footnote{http://www-hpcc.astro.washington.edu/tools/skid.html} with a linking length of $20~\mathrm{kpc} = 4~\epsilon$ and a density and number cut so that only structures that are virialised and which are represented by at least 100 particles are accepted. In Fig. \ref{fig:virgo}, we plot the mass function $n(M)$ (number of substructures of mass $M$) as a function of substructure mass $M$ for output at redshifts $z=5$ and $z=0$. There is no substantial difference between the mass functions for the different time-stepping schemes. 

For these low resolution runs we do not expect to see a significant difference between the three runs since the scale at which the standard scheme begins to take an insufficient number of time-steps corresponds approximately to the resolution scale of this simulation. This is just a confirmation that the dynamical time-stepping scheme also works for the extreme dynamics of a cosmological structure formation run. 

\subsection{Dependence on parameters} \label{chap:paramdep}

\subsubsection{Softening length $\epsilon$}

The standard time-step criterion (\ref{eq:TSstd}) depends directly on the artificial simulation parameter softening length $\epsilon$. There is however no physical basis for this definition. Furthermore, the functional form of the acceleration in centrally flat ($\gamma < 1$) haloes is problematic and can lead to nonsensical time-steps if the resolution is high enough (see Fig. \ref{fig:ts}). Even a simple 2-body problem is not treated properly by the standard time-stepping scheme, since the time-steps depend on acceleration which is not symmetric and again there is the meaningless dependence on the softening of the particles.

The dynamical time-stepping scheme only depends indirectly on the softening length. If two particles are close enough such that their forces are softened, we also use the softened values for the $\rho_{\mathrm{enc}}$. In this way the scheme also determines an appropriate dynamical time-step when the Green's function deviates from the Newtonian $1/r$.
Furthermore, the new dynamical time-stepping scheme may be used without modification in simulations where the softening is set to zero, i.e., where the interactions are never softened.

\subsubsection{Force opening angle $\theta$}

The opening angle $\theta$ determines the weighting of directly calculated forces to force contributions coming from the multipole expansion. This parameter mainly determines the accuracy of the force. By including the terms in the particle-bucket list, the dynamical scheme does not show a significant dependence on the choice of the force opening angle $\theta$.

\subsubsection{Cone viewing angle $\alpha$}

We normalised the viewing angle $\alpha$ so that the volume of the sphere and the cone in Fig. \ref{fig:cone} are equal. We also tried larger values of $\cos(\alpha) > 0.75$ (i.e. smaller angles) but the resulting time-step distribution did not follow the theoretical curves as well as for the case of $\cos(\alpha) = 0.75$ (especially close to the centre).

\subsubsection{Number of maxima}

Ideally, one would scan the particle's whole sky for the gravitationally dominating structure. But this would be computationally very expensive. With our choice of the 0.5 percentile largest $\rho_{\mathrm{enc}}$ cells in each of the cell and particle-bucket lists we find a good compromise between getting the correct dominating structure (i.e. low scattering of the enclosed density values) and computational speed. Having to consider multiple maxima is the main factor which makes the dynamical time-stepping scheme more expensive than the simple schemes used to-date. However, if we loosen the strict geometrical definition of the viewing cone, then faster schemes which rely on the hierarchical tree structure when scanning the sky for maxima and their surrounding mass become realisable. Such algorithmic improvements are being investigated and will be discussed in future work.

\subsubsection{Prefactor $\eta_{\mathrm{D}}$}

\cite{2001PhDT........21S} performed stability tests for a leapfrog scheme in the drift-kick-drift mode. The result was that 2-body orbits became unstable for choices of $\eta_{\mathrm{D}} \geq 0.1$. For these tests, the choice of time-steps was also based on the dynamical time of the 2-body problem.

We performed similar tests with the kick-drift-kick leapfrog scheme and found that $e = 0.9$ orbits become unstable in the mean field regime (i.e. without eccentricity correction) for choices of $\eta_{\mathrm{D}}$ too large. Of course by choosing a smaller value of $\eta_{\mathrm{D}}$, one always gets better precision but the computational costs become larger. With the choice of $\eta_{\mathrm{D}} = 0.03$ we found a compromise between stability and computational costs.

\subsection{Efficiency}

In order to quantify the efficiency of the dynamical time-stepping criterion in comparison with the standard criterion, we can compare the number of force evaluations for a given problem. In a spherically symmetric halo, the number of particles in a shell at radius $r$ with thickness $dr$ is given by
\begin{equation}
dN_{\mathrm{P}} = \frac{4 \pi \rho(r) r^2 dr}{\Mvir/\Nvir}~.
\end{equation}
The number of time-steps per time interval $\tau$ for each of these $dN_{\mathrm{P}}$ particles at radius $r$ is given by
\begin{equation}
N_{\mathrm{D}} = \frac{\tau}{\eta_{\mathrm{D}}} \sqrt{G \rho_{\mathrm{enc}}(r)} = \frac{\tau}{\eta_{\mathrm{D}}} \sqrt{\frac{G M(r)}{r^3}}
\end{equation}
in the dynamical case. In the case of the standard time-stepping scheme, this is instead given by
\begin{equation}
N_{\mathrm{S}} = \frac{\tau}{\eta_{\mathrm{S}}} \sqrt{\frac{|a(r)|}{\epsilon}} = \frac{\tau}{\eta_{\mathrm{S}}} \sqrt{\frac{G M(r)}{\epsilon~r^2}}~.
\end{equation}
We do not account for the actual block time-stepping scheme used in \textsc{pkdgrav} for this numerical estimation.

The number of force evaluations in that infinitesimal thin shell is now simply given by
\begin{equation}
dF_{\mathrm{S}} = dN_{\mathrm{P}} ~ N_{\mathrm{S}}
\end{equation}
respectively
\begin{equation}
dF_{\mathrm{D}} = dN_{\mathrm{P}} ~ N_{\mathrm{D}}~.
\end{equation}
The ratio
\begin{equation} \label{eq:workratio}
R_{\mathrm{E}} \equiv \frac{dF_{\mathrm{S}}}{dF_{\mathrm{D}}} = \frac{\eta_{\mathrm{D}}}{\eta_{\mathrm{S}}} \sqrt{\frac{r}{\epsilon}}~,
\end{equation}
shows that above $r_{\mathrm{eq}}$ defined by equation (\ref{eq:req}), i.e. the radius where both time-stepping criteria give the same value, the number of force evaluations at a given radius is always a factor $R_{\mathrm{E}} \propto \sqrt{r}$ larger.

\begin{figure}
	\centering
		\includegraphics[width=\columnwidth]{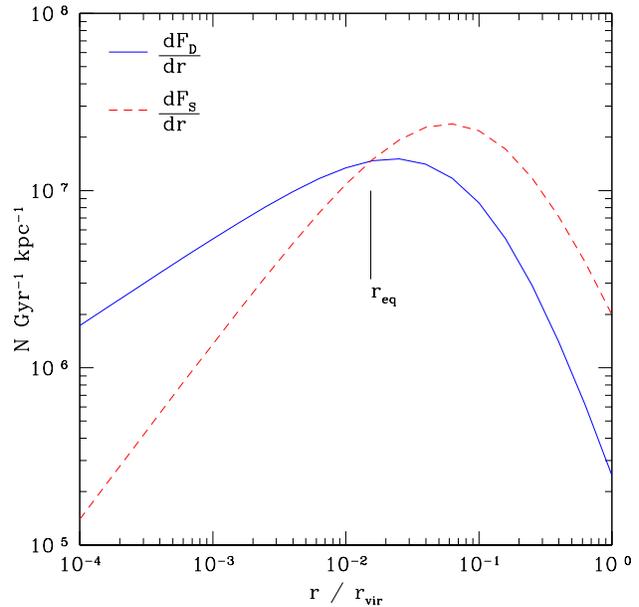}
	\caption{Number of force evaluations $N$ per Gyr per radius for the dynamical scheme $dF_{\mathrm{D}}/dr$ respectively $dF_{\mathrm{S}}/dr$ for the standard scheme for an NFW profile dark matter halo. For infinitesimal thin shells at radii larger than $r_{\mathrm{eq}}$, the standard time-stepping scheme has always a factor $R_{\mathrm{E}}$ (given by equation (\ref{eq:workratio})) more force evaluations per Gyr. In an NFW profile, the inner slope is $\gamma = 1$ and we have therefore the following asymptotic behaviour in the centre: $dF_{\mathrm{D}}/dr \propto r^{\frac{1}{2}}$ and $dF_{\mathrm{S}}/dr \propto r^1$.}
	\label{fig:work}
\end{figure}

In Fig. \ref{fig:work}, we plot the number of force evaluations per Gyr per radius $dF_{\mathrm{D}}/dr$ and $dF_{\mathrm{S}}/dr$ for the NFW profile dark matter halo used also in Fig. \ref{fig:ts} with $\Nvir = 7.5 \times 10^6$. The figure shows the asymptotic behaviour of the curves in the central region given by
\begin{equation}
\frac{dF_{\mathrm{D}}}{dr} \propto r^{2-\frac{3}{2}\gamma} \stackrel{\gamma = 1}{\propto} r^{\frac{1}{2}}
\end{equation}
respectively
\begin{equation}
\frac{dF_{\mathrm{S}}}{dr} \propto r^{\frac{5}{2}-\frac{3}{2}\gamma} \stackrel{\gamma = 1}{\propto} r^1.
\end{equation}
In the inner region, i.e at distances smaller than $r_{\mathrm{eq}}$ from the centre, more force evaluations are done in the dynamical time-stepping scheme, as expected. Here the standard scheme does not give small enough time-steps to follow the dynamics. Since in most cases of low resolution simulations $r_{\mathrm{eq}}$ (due to a clever choice of softening or $\eta_{\mathrm{S}}$) is around the resolution scale, the error is typically small. High resolution simulations can however become very slow if one wants to resolve the central region correctly with the standard time-stepping scheme.

In order to illustrate the efficiency gain, we calculate the number of force evaluations per Gyr for three different NFW haloes that have the same profile and specifications as the one used in Figs \ref{fig:ts} and \ref{fig:work}. We only changed the number of particles: they are $\Nvir = 7.5 \times 10^6, 7.5 \times 10^8$ and $7.5 \times 10^{10}$ particles within the virial radius. For the first halo we chose a softening length of $\epsilon_1 = 100~\mathrm{pc} \approx 3.5 \times 10^{-4}~\rvir$ and we scaled the softening of the other haloes according to the scaling of $r_{\mathrm{imp}}$ given by the solution of
\begin{equation}
r_{\mathrm{imp}} = h(r_{\mathrm{imp}})~,
\end{equation}
where $h(r)$ is the mean particle separation defined by
\begin{equation}
h(r) \equiv \sqrt[3]{\frac{\Mvir/\Nvir}{\rho(r)}}~.
\end{equation}
In other words, $r_{\mathrm{imp}}$ is the distance of the innermost particle to the geometrical centre of the halo and scales as
\begin{equation}
r_{\mathrm{imp}} \propto \sqrt[3-\gamma]{\frac{1}{\Nvir}} \stackrel{\gamma = 1}{\propto} \sqrt{\frac{1}{\Nvir}}
\end{equation}
resulting in $\epsilon_2 = 10~\mathrm{pc}$ and $\epsilon_3 = 1~\mathrm{pc}$ for the other softenings.

\begin{table}
	\caption{Number of force evaluations per Gyr for an NFW halo with different resolutions $\Nvir$.}
	\label{tab:efficiency}
	\centering
		\begin{tabular}{c c c c }
			\hline
			$\Nvir$ & $7.5 \times 10^6$   & $7.5 \times 10^8$      & $7.5 \times 10^{10}$\\
			\hline
			$F_{\mathrm{D}}$   & $7.612 \times 10^8$ & $7.612 \times 10^{10}$ & $7.612 \times 10^{12}$ \\
			$F_{\mathrm{S}}$   & $2.336 \times 10^9$ & $7.388 \times 10^{11}$ & $2.336 \times 10^{14}$ \\
			$\frac{F_{\mathrm{S}}}{F_{\mathrm{D}}}$ & 3.069 & 9.705 & 30.69 \\
			\hline
		\end{tabular}
\end{table}

The number of force evaluations per Gyr ($\tau$ = 1 Gyr) is given by
\begin{equation}
F_{\mathrm{D}} = \int_{r_{\mathrm{imp}}}^{\rvir} dF_{\mathrm{D}} \quad\mathrm{respectively}\quad 
F_{\mathrm{S}} = \int_{r_{\mathrm{imp}}}^{\rvir} dF_{\mathrm{S}}~,
\end{equation}
and the numerical results can be found in Table \ref{tab:efficiency}. We see that $F_{\mathrm{D}}$ scales as $F_{\mathrm{D}} \propto \Nvir$ whereas $F_{\mathrm{S}}$ scales approximately as $F_{\mathrm{S}} \propto \Nvir^{1.25}$ in the case of an NFW profile halo. This specific scaling of $F_{\mathrm{S}}$ is due to the scaling of the softening length $\epsilon \propto r_{\mathrm{imp}}$ resulting in the general scaling of
\begin{equation}
F_{\mathrm{S}} \propto \Nvir^{1+\frac{1}{2(3-\gamma)}}~.
\end{equation}
For a different scaling of the softening, one gets of course a different scaling of $F_{\mathrm{S}}$, e.g. $\epsilon \propto \Nvir^{-1/3}$ results in $F_{\mathrm{S}} \propto \Nvir^{7/6}$ independent of $\gamma$. This again shows the strong dependence of the standard scheme on the softening length $\epsilon$ and that the the dynamical time-stepping scheme is much more efficient than the standard scheme for high resolution simulations.

\section{Conclusions}\label{chap:conclusions}

We have developed a physically motivated time-stepping scheme that is based on the true dynamical time of the particle. 
We also derive an eccentricity correction for a general leapfrog integration scheme. The combination of these schemes allows us to follow quite general dynamical systems that may contain a mixture of collisionless and collisional interacting components. Compared to the standard time-stepping scheme used in many $N$-body codes it has the following advantages:

\begin{enumerate}
\item It does not depend directly on ad-hoc parameters such as the softening length $\epsilon$.
\item It gives physically correct time-steps in dark matter haloes with arbitrary central cusp slopes.
\item It is faster in high resolution simulations.
\item It allows orbits with eccentricity $e\rightarrow1$ to be followed correctly.
\item It allows us to follow complex dynamical systems where scattering events may be important.
\end{enumerate}

The main conclusion is that one should use a time-step criterion that is based on the dynamical time. This scheme shows the optimum scaling with the number of particles and always gives a physically motivated time-step.

\section*{Acknowledgment}

It is a pleasure to thank J\"urg Diemand, Andrea Macci\`o and Prasenjit Saha for helpful comments. All the simulations were carried out at zBox1\footnote{http://www.zbox1.org} and zBox2\footnote{http://www.zbox2.org} supercomputers at University Zurich. It is also a special pleasure to thank Doug Potter for his immense work with getting the zBox2 working.

\appendix

\section{Hamiltonian formalism} \label{chap:hamilton}

For a physical system that is described by the state $\vec{z} = (\vec{q},\vec{p})$, where $\vec{q}$ is the coordinate and $\vec{p}$ the conjugate momentum vector, and this system is evolved under a Hamiltonian $H$, we can write the formal time evolution (Hamilton equations) as \citep{1994AJ....108.1962S,1993CeMDA..56...27Y}
\begin{equation}
\frac{d\vec{z}}{dt} = \left \{ \vec{z},H \right \} 
\end{equation}
where $\left \{ , \right \}$ denote the Poisson brackets defined by
\begin{equation}
\left \{ g,h \right \} \equiv \sum_i^f \left( 
\frac{\partial g}{\partial q_i} \frac{\partial h}{\partial p_i} -
\frac{\partial g}{\partial p_i} \frac{\partial h}{\partial q_i} \right)~.
\end{equation}
We can define the operator
\begin{equation}
\hat{H} \vec{z} \equiv \left \{ \vec{z},H \right \}
\end{equation}
and write down a formal solution to the time evolution
\begin{equation}
\vec{z}(t) = e^{t \hat{H}} \vec{z}_0~.
\end{equation}

In computer simulations, the system is evolved by using a specific scheme in order to update positions and velocities. In \textsc{pkdgrav}, the following leapfrog scheme is used: during a time-step $\Delta T$, first the velocities are updated (kick mode) with a time-step of $\frac{\Delta T}{2}$ then the new positions are calculated (drift mode) using the new velocities with a time-step of $\Delta T$ and finally the velocities are updated to the final values at $\Delta T$ with again a half-step of $\frac{\Delta T}{2}$. This is therefore called the kick-drift-kick mode and can be described by
\begin{equation}
\vec{z}(\Delta T) = e^{\frac{\Delta T}{2} \hat{H}_{\mathrm{K}}} e^{\Delta T \hat{H}_{\mathrm{D}}} e^{\frac{\Delta T}{2} \hat{H}_{\mathrm{K}}} \vec{z}_0
\end{equation}
where we have split the true Hamiltonian into a drift and a kick part
\begin{equation}
\hat{H} = \hat{H}_{\mathrm{D}} + \hat{H}_{\mathrm{K}}~.
\end{equation}
By using the Baker-Campbell-Hausdorff series, where we can calculate the higher order terms by an elegant method developed by \citet{2000JMP....41.2434R}, we can write the approximate operator $\hat{H}_{A}$ under which the system is evolved
\begin{equation}
\vec{z}(\Delta T) = e^{\frac{\Delta T}{2} \hat{H}_{\mathrm{K}}} e^{\Delta T \hat{H}_{\mathrm{D}}} e^{\frac{\Delta T}{2} \hat{H}_{\mathrm{K}}} \vec{z}_0 = e^{\Delta T \hat{H}_{\mathrm{A}}} \vec{z}_0
\end{equation}
by
\begin{equation}
\hat{H}_{\mathrm{A}} = \hat{H}_0 + \Delta T^2 \hat{H}_2 + \Delta T^4 \hat{H}_4 + O(\Delta T^6)
\end{equation}
where 
\begin{eqnarray}
\hat{H}_0 &=& \hat{H}_{\mathrm{D}} + \hat{H}_{\mathrm{K}} = \hat{H} \\
\hat{H}_2 &=& \frac{1}{12} \left[\left[\hat{H}_{\mathrm{K}},\hat{H}_{\mathrm{D}}\right],\hat{H}_{\mathrm{D}}\right]
- \frac{1}{24} \left[\left[\hat{H}_{\mathrm{D}},\hat{H}_{\mathrm{K}}\right],\hat{H}_{\mathrm{K}}\right] \\
\hat{H}_4 &=& \frac{7}{5760} \left[\left[\left[\left[\hat{H}_{\mathrm{D}},\hat{H}_{\mathrm{K}}\right],\hat{H}_{\mathrm{K}}\right],\hat{H}_{\mathrm{K}}\right],\hat{H}_{\mathrm{K}}\right] \\
&& \frac{7}{1440} \left[\left[\left[\left[\hat{H}_{\mathrm{D}},\hat{H}_{\mathrm{K}}\right],\hat{H}_{\mathrm{D}}\right],\hat{H}_{\mathrm{K}}\right],\hat{H}_{\mathrm{K}}\right] \nonumber \\
&& -\frac{1}{360} \left[\left[\left[\hat{H}_{\mathrm{D}},\hat{H}_{\mathrm{K}}\right],\hat{H}_{\mathrm{K}}\right],\left[\hat{H}_{\mathrm{D}},\hat{H}_{\mathrm{K}}\right]\right] \nonumber \\
&& -\frac{1}{180} \left[\left[\left[\left[\hat{H}_{\mathrm{K}},\hat{H}_{\mathrm{D}}\right],\hat{H}_{\mathrm{K}}\right],\hat{H}_{\mathrm{D}}\right],\hat{H}_{\mathrm{D}}\right] \nonumber \\
&& -\frac{1}{360} \left[\left[\left[\hat{H}_{\mathrm{K}},\hat{H}_{\mathrm{D}}\right],\hat{H}_{\mathrm{D}}\right],\left[\hat{H}_{\mathrm{K}},\hat{H}_{\mathrm{D}}\right]\right] \nonumber \\
&& -\frac{1}{720} \left[\left[\left[\left[\hat{H}_{\mathrm{K}},\hat{H}_{\mathrm{D}}\right],\hat{H}_{\mathrm{D}}\right],\hat{H}_{\mathrm{D}}\right],\hat{H}_{\mathrm{D}}\right] \nonumber~.
\end{eqnarray}
Here, $[ , ]$ denote the commutator brackets defined by
\begin{equation}
[A,B] = AB - BA
\end{equation}
for non-commutative operators $A$ and $B$. By using the definitions of the operators and the Jacobi identity for Poisson brackets, we can calculate the approximate Hamiltonian
\begin{equation}
H_{\mathrm{A}} = H_0 + \Delta T^2 H_2 + \Delta T^4 H_4 + O(\Delta T^6)
\end{equation} 
where 
\begin{eqnarray}
H_0 &=& H_{\mathrm{D}} + H_{\mathrm{K}} = H \\
H_2 &=& \frac{1}{12} \left\{\left\{H_{\mathrm{K}},H_{\mathrm{D}}\right\},H_{\mathrm{D}}\right\}
- \frac{1}{24} \left\{\left\{H_{\mathrm{D}},H_{\mathrm{K}}\right\},H_{\mathrm{K}}\right\} \\
H_4 &=& \frac{7}{5760} \left\{\left\{\left\{\left\{H_{\mathrm{D}},H_{\mathrm{K}}\right\},H_{\mathrm{K}}\right\},H_{\mathrm{K}}\right\},H_{\mathrm{K}}\right\} \\
&& \frac{7}{1440} \left\{\left\{\left\{\left\{H_{\mathrm{D}},H_{\mathrm{K}}\right\},H_{\mathrm{D}}\right\},H_{\mathrm{K}}\right\},H_{\mathrm{K}}\right\} \nonumber \\
&& -\frac{1}{360} \left\{\left\{\left\{H_{\mathrm{D}},H_{\mathrm{K}}\right\},H_{\mathrm{K}}\right\},\left\{H_{\mathrm{D}},H_{\mathrm{K}}\right\}\right\} \nonumber \\
&& -\frac{1}{180} \left\{\left\{\left\{\left\{H_{\mathrm{K}},H_{\mathrm{D}}\right\},H_{\mathrm{K}}\right\},H_{\mathrm{D}}\right\},H_{\mathrm{D}}\right\} \nonumber \\
&& -\frac{1}{360} \left\{\left\{\left\{H_{\mathrm{K}},H_{\mathrm{D}}\right\},H_{\mathrm{D}}\right\},\left\{H_{\mathrm{K}},H_{\mathrm{D}}\right\}\right\} \nonumber \\
&& -\frac{1}{720} \left\{\left\{\left\{\left\{H_{\mathrm{K}},H_{\mathrm{D}}\right\},H_{\mathrm{D}}\right\},H_{\mathrm{D}}\right\},H_{\mathrm{D}}\right\} \nonumber~.
\end{eqnarray}
Note that the replacement of the commutator brackets by the Poisson brackets is not trivial. The term $H_4$ is not used in our method. However, since $H_4$ is not explicitly given in the recent literature we present it here for completeness.

\bibliography{RDB_S}

\label{lastpage}

\end{document}